\documentclass[12pt,a4paper]{article} 
\pdfoutput=1

\newcommand{\blind}{0}
 
\usepackage{amsfonts}
\usepackage{amsmath}
\usepackage{amssymb}
\usepackage{amsthm}
\usepackage{authblk}
\usepackage{booktabs}
\usepackage[margin=5pt,font=small,labelfont=bf,labelsep=colon, skip=2pt, parskip=5pt]{caption}
\usepackage[usenames]{color}
\usepackage{epsfig}
\usepackage{enumerate}
\usepackage{float}
\usepackage{graphicx}
\usepackage{indentfirst}
\usepackage{latexsym}
\usepackage{lscape}
\usepackage{multirow}
\usepackage{multicol}
\usepackage{microtype}
\usepackage{rotate}
\usepackage{subfig}
\usepackage{tabularx}
\usepackage{verbatim}

\usepackage{multirow}
\usepackage{natbib}

\newlength{\defaultbaselineskip}
\begin{document}

\begin{titlepage}
%\title{\bf A Dynamic Conditional Approach for Forecasting Optimal Weights in Portfolio Allocations\setcounter{footnote}{0}}
\title{\bf A Dynamic Conditional Approach to Portfolio Weights Forecasting\setcounter{footnote}{0}}

\if0\blind
{
  \author{
    Fabrizio Cipollini\\
    DiSIA, Universit\`a di Firenze, Italy \\
    e-mail: \texttt{fabrizio.cipollini@unifi.it}
    \vspace{1em}\\
    Giampiero M.~Gallo\thanks{Address correspondence to: Giampiero M. Gallo,  Corte dei conti, Sezione regionale di controllo per la Lombardia, Via Marina 5, 20121 Milan, Italy, Email: \texttt{giampiero.gallo@nyu.edu}. The views expressed in the article are those of the authors and do not involve the responsibility of the Corte dei conti.}
    Corte dei conti and NYU in Florence, Italy \\
    e-mail: \texttt{giampiero.gallo@nyu.edu}
    \vspace{1em}\\
    Alessandro Palandri\\
    DiSIA, Universit\`a di Firenze, Italy \\
    and\\
    DCU Business School, Dublin City University\\    
    e-mail: \texttt{alessandro.palandri@gmail.com}
    }
} \fi
\if1\blind
{
  \author{}
  \bigskip
  \bigskip
  \bigskip
  \medskip
} \fi

% \author{Fabrizio Cipollini$^{1}$, Giampiero M. Gallo$^{2,}$\footnote{Address correspondence to: Giampiero M. Gallo,  Corte dei conti, Sezione regionale di controllo per la Lombardia, Via Marina 5, 20121 Milan, Italy, Email: {\tt giampiero.gallo@nyu.edu}. The views expressed in the article are those of the authors and do not involve the responsibility of the Corte dei conti.} ~and Alessandro Palandri$^{1,3}$}
% \affil{$^{1}$Dipartimento di Statistica, Informatica, Applicazioni (DiSIA)\\ {\it G. Parenti}, Universit\`a degli Studi di Firenze}
% \affil{$^{2}$Italian Court of Audits, and New York University Florence}
% \affil{$^{3}$DCU Business School, Dublin City University}
\date{This Version: \today}
\end{titlepage}

\maketitle{}
% \thispagestyle{empty}

%\begin{center}
%\begin{Large}
%{\tt DRAFT PAPER - PLEASE DO NOT CIRCULATE}
%\end{Large}
%\end{center}

\renewcommand{\arraystretch}{1.5}
\setlength{\baselineskip}{6.5mm} %\vspace{1.5cm}
\begin{abstract}
We build the time series of optimal realized portfolio weights from high-frequency data and we suggest a novel Dynamic Conditional Weights ($\mathsf{DCW}$) model for their dynamics. $\mathsf{DCW}$ is benchmarked against popular {\it model-based} and {\it model-free} specifications in terms of weights forecasts and portfolio allocations. Next to portfolio variance, certainty equivalent and turnover, we introduce the break-even transaction costs as an additional measure that identifies the range of transaction costs for which one allocation is preferred to another. By comparing minimum-variance portfolios built on the components of the Dow Jones 30 Index, the proposed $\mathsf{DCW}$ overall attains the best allocations with respect to the measures considered, for any degree of risk-aversion, transaction costs and exposure.
%The paper proposes the Dynamic Conditional Weights ($\mathsf{DCW}$), a new approach modeling directly the time series of optimal portfolio weights from high-frequency data. $\mathsf{DCW}$ is benchmarked against representative {\it model-based} and {\it model-free} specifications in terms of weights forecasts and portfolio allocations. Next to portfolio variance, certainty equivalent and turnover, we introduce the break-even transaction costs as a novel measure which identifies the range of transaction costs giving rise to the preference of one allocation over another. By comparing minimum-variance portfolios built on the components of the Dow Jones 30 Index, we find that the proposed $\mathsf{DCW}$ overall attains the best allocations with respect to the measures considered, for any degree of risk-aversion, transaction costs and exposure.
\end{abstract}
\vspace{2cm}
\noindent{\bf Keywords:} Portfolio Allocation, Realized Volatility, Realized Correlations, Dynamic Conditional Modeling, Portfolio Weights Modeling.
\textcolor{white}{space}\\
\noindent{\bf JEL classification:} C32, C53, G11, G17.

\pagebreak
\setcounter{page}{1}

\section{Introduction \label{intro}}

A key aspect of active portfolio management is forecasting the weights that optimize portfolio holdings with respect to some  representative measure of the investor's preferences. Since \citet{Markowitz:1952}, these forecasts of the optimal portfolio weights are generally derived from the forecasts of conditional moments of asset returns. The availability of realized measures from high-frequency data allows for {\it model-based} and {\it model-free} forecasting of conditional variance-covariance (var-cov) matrices and, by successive manipulation thereof, of optimal portfolio weights: see, e.g.,  \citet{Ait-Sahalia:Fan:Xiu:2010}, \citet{Christensen:Kinnerbrock:Podolskij:2010}, \citet{BarndorffNielsen:Hansen:Lunde:Shephard:2011}, \citet{Zhang:2011} and \citet{Bibinger:Hautsch:Malec:Reiss:2014}, among others. 

%Modeling and forecasting conditional variance-covariance (var-cov) matrices of asset returns is an essential element of active portfolio management since \citet{Markowitz:1952}. The availability of high-frequency data and the ensuing realized measures have originated a series of developments to the {\it model-based} and {\it model-free} forecasting of conditional second moments and, in turn, of optimal portfolio weights: see, e.g.,  \citet{Ait-Sahalia:Fan:Xiu:2010}, \citet{Christensen:Kinnerbrock:Podolskij:2010}, \citet{BarndorffNielsen:Hansen:Lunde:Shephard:2011}, \citet{Zhang:2011} and \citet{Bibinger:Hautsch:Malec:Reiss:2014}, among others. Nonetheless, despite modeling differences, existing approaches share the indirect inclusion of the information content of realized measures into the forecasts of optimal portfolio weights via forecasts of the conditional var-cov.

The {\it model-based} approaches are inspired by the logic behind  Multivariate-GARCH (MGARCH) models\footnote{For a review of MGARCH models see \citet{Bauwens:Laurent:Rombouts:2006}.} with the substantial difference that information is extracted from realized measures rather than low-frequency estimates of the second moments, such as the outer product of the vector of returns (or their residuals after some filtration). Examples of these approaches are the fractionally integrated processes of \citet{Chiriac:Voev:2011}, the vector autoregressions of \citet{Callot:Kock:Medeiros:2017} and the specifications based on the Wishart distribution of \citet{Gourieroux:Jasiak:Sufana:2009}, \citet{Golosnoy:2012}, \citet{Noureldin:Shephard:Sheppard:2012} and \citet{Jin:Maheu:2013}, among others. Within this framework it is not uncommon to separately model conditional variances\footnote{Amongst the various approaches to volatility modeling that make use of realized measures are the Heterogeneous Autoregressive model (HAR) of \citet{Corsi:2009} and \citet{Corsi:Audrino:Reno:2012}, the Multiplicative Error Model (MEM) of \citet{Engle:2002} and \citet{Brownlees:Cipollini:Gallo:2012} and the HEAVY of \citet{Shephard:Sheppard:2010}, as a particular case of the vector-MEM of \citet{Cipollini:Engle:Gallo:2013}. For a survey see \citet{Andersen:Bollerslev:Christoffersen:Diebold:2006} and \citet{Park:Linton:2012}, among others.} and correlation matrices to achieve a good balance between parameter parsimony and richness in the description of the second order dynamics.

On the other hand, {\it model-free} approaches, also referred to as {\it nonparametric}, impose driftless random-walk dynamics to the conditional second moments, and thus eliminate the parameter estimation problem altogether. However, for large cross-sectional dimensions, the lag-1 realized var-cov matrices may result in  extreme portfolio weights, poor portfolio performance out-of-sample (OOS) and even positive-semidefiniteness (psd). To mitigate this problem, various shrinkage approaches are available: the most direct is to impose constraints on the portfolio weights\footnote{While the target is usually defined by portfolio weights with no short-sale (positivity) constraints, other alternatives are also possible, i.e. equal weights.} as in \citet{Jagannathan:Ma:2003}, \citet{ElKaroui:2010}, \citet{Fan:Zhang:Yu:2012} and \citet{Gandy:Veraart:2013}. Shrinkage of the realized var-cov matrix has been proposed by \citet{Fan:Fan:Lv:2008}, \citet{Fan:Liao:Mincheva:2011}, \citet{Ledoit:Wolf:2012}, \citet{Tao:Wang:Yao:Zou:2011}, \citet{Tao:Wang:Zou:2013}, \citet{Fan:Furger:Xiu:2016} and \citet{Ait-Sahalia:Xiu:2017}, to name a few. Ideas behind these approaches may also be traced back to the MGARCH literature and consist of imposing a factor structure to the returns and a sparse error var-cov matrix with blocks defined by some characteristics of the assets such as sector, industry, etc.

In this paper we introduce the Dynamic Conditional Weights ($\mathsf{DCW}$), an approach which emerges when expressing the autoregressive representation of the portfolio-variance optimization problem in terms of a time-independent weighting matrix. The result is a specification in which the forecast of the conditional portfolio weights derives from a linear function of past conditional weights and past realized (hence observable) weights. When associated with suitable estimation procedures, the main advantage of $\mathsf{DCW}$ with respect to standard {\it model-based} approaches is the circumvention of the curse of dimensionality problem. With respect to the {\it model-free} approaches, $\mathsf{DCW}$ does not require the imposition of particular var-cov matrix structures nor discretionary choices about the level of shrinkage. 

Focusing on the minimum-variance allocation, empirical results show that $\mathsf{DCW}$ outperforms {\it model-based} and {\it model-free} approaches in terms of out-of-sample portfolio variance, certainty equivalence and turnover \citep{DeMiguel:Garlappi:Uppal:2009}.  Since transaction costs may significantly alter the outlook in the performance of the approaches, we introduce the {\it Break-Even Transaction Costs} as a more comprehensive measure of forecasting performance confirming the goodness of the $\mathsf{DCW}$ allocation in terms of minimal portfolio variance and turnover. Furthermore, since the {\it model-based} and {\it model-free} literatures have proceeded on somewhat parallel tracks,\footnote{For example, the most recent contributions to the {\it model-free} literature, such as \citet{Fan:Furger:Xiu:2016} and \citet{Ait-Sahalia:Xiu:2017}, do not benchmark their approaches to any of the {\it model-based}.} a contribution of this paper is a comparison across approaches, assessing the quality of the respective forecasts and portfolio allocations. 

The paper is organized as follows. Section \ref{portfolio} introduces the optimal portfolio allocation problem. The {\it model-based} and {\it model-free} approaches are discussed in Section \ref{proj_cov}. Section \ref{weights} introduces the direct modeling of the portfolio weights. Measures of performance, data and results are presented in Section \ref{results}. Section \ref{concl} concludes.

\section{Minimum Variance Portfolio \label{portfolio}}

Following \citet{Ait-Sahalia:Xiu:2017}, \citet{Fan:Furger:Xiu:2016}, \citet{Behr:Guettler:Miebs:2013} and \citet{Fan:Zhang:Yu:2012}, \citet{Fan:Fan:Lv:2008}, among others,\footnote{Other examples are \citet{Bednarek:Patel:2018}, \citet{Maillet:Tokpavi:Vaucher:2015}, \citet{Candelon:Hurlin:Tokpavi:2012}, \citet{Scherer:2011}, \citet{Clarke:deSilva:Thorley:2008}, etc.} we focus on minimizing portfolio variance, which allows for a clean evaluation of the contribution of modeling and forecasting second moments to the optimal allocation. Furthermore, the minimum-variance portfolio has often been found to perform equally well as, if not better than, the mean-variance portfolio, even when measured in terms of Sharpe ratios: see \citet{DeMiguel:Garlappi:Uppal:2009} and \citet{DeMiguel:Martin-Utrera:Nogales:2014}. 

Letting $\Omega_{t}$ be the time $(t-1)$--conditional variance-covariance matrix of the $(M\times 1)$ vector of returns $r_{t}$ in excess of the risk-free rate, the optimal {\it relative} weights that minimize portfolio variance are given by:
\begin{equation}\label{eq:portfolio_weights}
\omega_{t} = \frac{\textcolor{white}{\iota'}\Omega_{t}^{-1}\iota}{\iota'\Omega_{t}^{-1}\iota}
\end{equation}
where $\iota$ is the $(M\times 1)$ unit vector. Such weights are optimal for investors maximizing the following quadratic utility:
\begin{equation}\label{eq:portfolio_utility}
V_{t} = -\frac{\gamma}{2}\omega_{t}'\Omega_{t}\omega_{t} \quad \text{ s.t. } \quad \iota'\omega_{t}=1
\end{equation}
where $\gamma$ is the investor's risk-aversion. Although inconsequential in the utility specification of equation (\ref{eq:portfolio_utility}), the level of risk-aversion $\gamma$ becomes relevant in the presence of transaction costs: see Section \ref{BETC}. While the {\it model-based} and the {\it model-free} literature have focused on generating forecasts of $\Omega_{t}$ to plug in equation (\ref{eq:portfolio_weights}), the proposed $\mathsf{DCW}$ will model and forecast $\omega_{t}$ directly.

Throughout the paper we consider the portfolio allocation problem of a {\it day trader} type of investor who closes positions at the end of each trading day. By so doing, we can ignore pre--market or after--hours exchanges which follow different price formation dynamics and for which high-frequency observations are not available. Moreover, it allows us to neatly bypass all potential problems arising from short positions that stretch over long periods of time.

\begin{comment}
\footnote{Notice that these weights are the limiting case of the optimal {\it relative} mean-variance weights when expected returns are assumed equal across assets. This is the case when expected returns are neither statistically nor economically different across assets or when they are shrunk toward the common mean. Specifically, letting $\mu_{t}$ be the conditional mean of the returns, the optimal {\it relative} mean-variance weights are given by $\omega_{t}=(\iota'\Omega_{t}^{-1}\mu_{t})^{-1}\cdot\Omega_{t}^{-1}\mu_{t}$. For $\mu_{t}\rightarrow m_{t}\iota$, where $m_{t}$ is the common mean, minimum-variance is the optimal mean-variance allocation.} 
\footnote{We have also adapted {\it model-based} specifications to modeling and forecasting of $\Omega_{t}^{-1}$ but the resulting allocations were no better than those attained from working with $\Omega_{t}$. Hence, we omit further analyses and reporting.}
\end{comment}

\section{Projecting Covariances \label{proj_cov}}

\subsection{Model-Based Approaches \label{model_based}}
Let $S_{t}$ be a realized measure of the var-cov matrix of $M$ assets at time $t$ and $\Omega_{t}\equiv\mathbb{E}_{t-1}[S_{t}]$ be its conditional expectation in $(t-1)$. {\it Model-based} approaches provide dynamic structures for $\Omega_{t}$ in terms of lags of $\Omega_{t}$ and $S_{t}$. In general, {\it model-based} approaches inspired by the most popular MGARCH models generate positive definite (pd) predictions $\Omega_{t}$ under the weakest condition that the realizations $S_{t-1}$ are psd of rank one\footnote{Predictions may fail to be pd for extremely large psd realizations for which $\Omega_{t}\propto S_{t-1}$.}. However, there is an inherent trade-off to the modeling of pd matrices: parsimony of model parameters {\it versus} richness in the description of the second order dynamics. In fact, the number of parameters to be jointly estimated is generally a power function of the cross-sectional dimension $M$. For example, in the Dynamic Conditional Correlations of \citet{Engle:2002b}, with {\it targeting}\footnote{Variance {\it targeting}, proposed by \citet{Engle:Mezrich:1996}, is the setting of the model's unconditional variance to its sample counterpart. In the multivariate case, {\it targeting} is particularly convenient as it eliminates $M(M+1)/2$ parameters from variance specifications and $M(M-1)/2$ parameters from correlation specifications.} the order is $M^{2}$, $M^{1}$ and $M^{0}$ for the full, diagonal and scalar matrices of parameters, respectively. Nevertheless, the dimensionality problem may be circumvented altogether by the element-by-element modeling of the conditional variance-covariance matrices in the order prescribed by the Sequential Conditional Correlations (SCC) decomposition of \citet{Palandri:2009}.

Representative specifications of the {\it model-based} class to be considered in the empirical analysis that follows are Volatility Timing ($\mathsf{VT}$) and Dynamic Conditional Correlations ($\mathsf{DCC}$) based on realized measures. While both approaches model the conditional variances of the returns, only $\mathsf{DCC}$ also models the conditional correlation matrix while $\mathsf{VT}$ sets it equal to the identity matrix. We model the $M$ conditional variances using the benchmark HAR specification of \citet{Corsi:2009}. Let $s_{i,t}^{2}$ be a realized measure of the variance of asset $i$ at time $t$ and $\sigma_{i,t}^{2}\equiv\mathbb{E}_{t-1}[s_{i,t}^{2}]$ its conditional expectation at $(t-1)$, then:
\begin{equation*}
\sigma_{i,t}^{2} = \alpha_{i,0} + \alpha_{i,1} \cdot  s_{i,t-1}^{2} + \alpha_{i,2}\cdot\frac{1}{5}\sum_{j=1}^{5}s_{i,t-j}^{2} + \alpha_{i,3}\cdot\frac{1}{22}\sum_{j=1}^{22}s_{i,t-j}^{2} 
\end{equation*}
which links the conditional variance $\sigma_{i,t}^{2}$ to past realizations over daily, weekly and monthly time intervals. 

In $\mathsf{DCC}$, the conditional var-cov matrix is decomposed into standard-deviation $D_{t}$ and correlation $R_{t}$ matrices: $\Omega_{t}=D_{t}R_{t}D_{t}$. The elements of $D_{t}$ are populated with the square-root of the HAR variances, while the elements of $R_{t}$ are modeled jointly using the Dynamic Conditional Correlation (1,1) specification\footnote{Although (1,1) specifications are well suited in the empirical applications, as highlighted by \citet{Hansen:Lunde:2005}, further lags may be considered.} with {\it targeting}:
\begin{equation}\label{eq:dcc}
R_{t} = \left(\overline{P}-A\overline{P}A'-B\overline{P}B'\right)+A P_{t-1}A'+BR_{t-1}B',
\end{equation}
where $\overline{P}$ is the sample average of the realized correlation matrices $P_{t}$ and $A$ and $B$ are either full, diagonal or scalar matrices of parameters. With a perspective on portfolios constructed over a vast number of assets, it should be noted that, with parameters of order $M^{0}$, scalar $\mathsf{DCC}$ with {\it targeting} is the only scalable specification of the three. We estimate model parameters both by least-squares and Gaussian quasi-maximum-likelihood\footnote{$\mathsf{DCC}$ estimated by quasi-maximum-likelihood is the HEAVY of \citet{Noureldin:Shephard:Sheppard:2012} without the overnight components.}. Finding that the former is orders of magnitude faster than the latter and delivers superior OOS results, we only report and discuss the findings pertaining to the least-squares estimation.

\subsection{Model-Free Approaches \label{model_free}}

Underlying the {\it model-free} approaches is the assumption that the realized var-cov matrices follow a driftless random-walk process from which $\Omega_{t}=S_{t-1}$. Although this assumption eliminates estimation and scalability problems, the literature on volatility (GARCH, MGARCH and Realized Variance models) has invariably shown how stationary specifications consistently outperform the random-walk, both in-sample (IS) and OOS. Furthermore, in contrast to the {\it model-based} specifications, problems do arise using the random-walk when $S_{t-1}$ is psd. With respect to this issue, \citet{Ait-Sahalia:Xiu:2017} adopt the following three approaches.

The first consists of aggregating daily realized measures into $k$-period  (for an arbitrary $k$, e.g. bimonthly) var-cov matrices, which delivers pd $k$-period measures unsuitable at the daily level. Furthermore, the exclusion of the overnight movements, not captured by the realized measures, while irrelevant from the perspective of a day trader, may accumulate undesirable effects over $k$-periods. 

The second is to express the psd $S_{t-1}$ as the sum of two rank-deficient matrices: the first arising from a factor structure\footnote{Factor decompositions have been studied extensively in the MGARCH literature and generated the Factor-GARCH family of models: see \citet{Diebold:Nerlove:1989}, \citet{Engle:Ng:Rothschild:1990}, \citet{Alexander:Chibumba:1997}, \citet{Sentana:1998}, among others. Although, due to the {\it curse of dimensionality}, Factor-GARCH models have not been particularly successful in dealing with both flexibility and feasibility, the new idea of the {\it model-free} literature is to decompose the realizations into factors and residual components to shrink.} of the data, as in \citet{Fan:Fan:Lv:2008}, \citet{Fan:Liao:Mincheva:2011}, \citet{Fan:Furger:Xiu:2016} and \citet{Ait-Sahalia:Xiu:2017}, and the second being the residual var-cov matrix. Calibrating the shrinkage of the latter toward a diagonal or block-diagonal structure\footnote{Block-diagonal structures based on characteristics such as sector, industry, etc. had already been investigated in the MGARCH literature: for example, see \citet{Billio:Caporin:Gobbo:2006} and \citet{Billio:Caporin:2009}.} allows to achieve pd of the recombined matrix. 

The third consists of controlling for portfolio exposure $EC$ (where $EC=1$ and $EC=\infty$ are the no short-selling, respectively, the unconstrained portfolios) by adding the constraint $\sum_{i=1}^{M}|w_{i,t}|\le EC$ to the optimization problem in equation (\ref{eq:portfolio_utility}). Proposed, among others, by \citet{Jagannathan:Ma:2003}, exposure constraints help reduce the effects of estimation and forecast errors, portfolio turnover and associated transaction costs. However, exposure constraints alone are not enough to turn the optimization problem in (\ref{eq:portfolio_utility}) from ill- to well-behaved when var-cov matrices are rank-deficient.

For what matters here, as a benchmark specification for the {\it model-free} class we adopt the raw realized $S_{t-1}$ with exposure constraints. In so doing, we rely on some findings - presented independently by \citet[Figure 5]{Fan:Furger:Xiu:2016}, and \citet[Figure 6]{Ait-Sahalia:Xiu:2017} - showing that factor structures and shrinkages do not bring about significant improvements in the OOS results. Following on their outcome that the optimal exposure is $EC=2$, we investigate exposure constraints between $1$ and $2$, i.e. $EC=\{1.00, 1.25, 1.50, 1.75, 2.00\}$, keeping the case of no constraints on the weights ($EC=\infty$) as a reference. In fact, as $EC$ increases, transaction costs become larger and larger; thus $EC>2$ is suboptimal as the resulting portfolios exhibit larger OOS variances and higher transaction costs. By the same token, values of $EC<2$ should not be discarded {\it a priori} as a larger OOS variance may be offset, in some measure, by lower transaction costs.

\section{Dynamic Conditional Weights Modeling \label{weights}}

\begin{figure}[p]
\captionsetup{singlelinecheck=off}
\caption[ac]{Realized portfolio weights over the entire 2005-2015 period for Apple (top-left), Boeing (top-right), Johnson \& Johnson (bottom-left) and Merck (bottom-right).\medskip}\label{fig:temp}
\hspace{0.2cm}\noindent\includegraphics[width=0.48\textwidth, height=0.12\textheight]{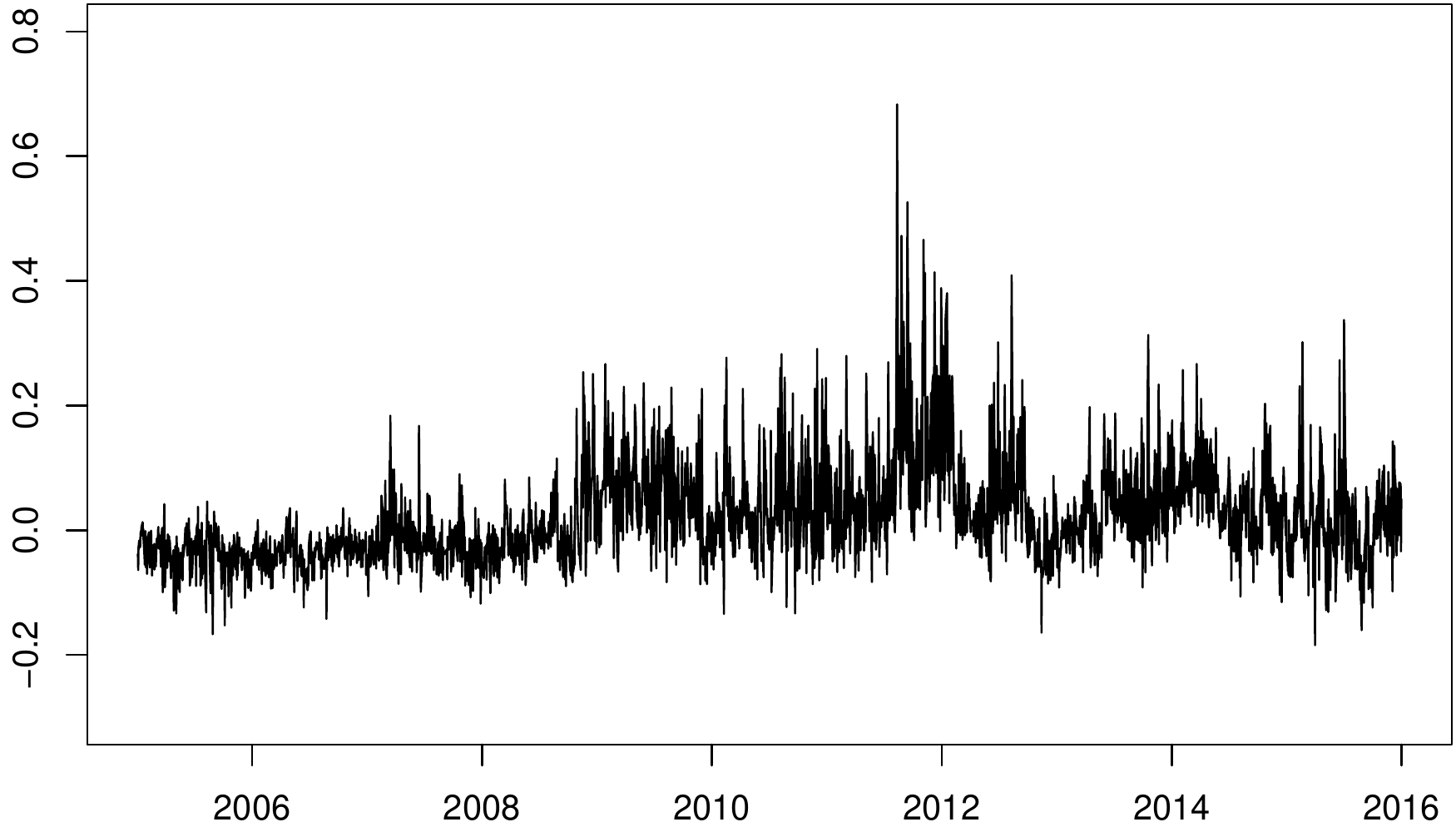}
\noindent\includegraphics[width=0.48\textwidth, height=0.12\textheight]{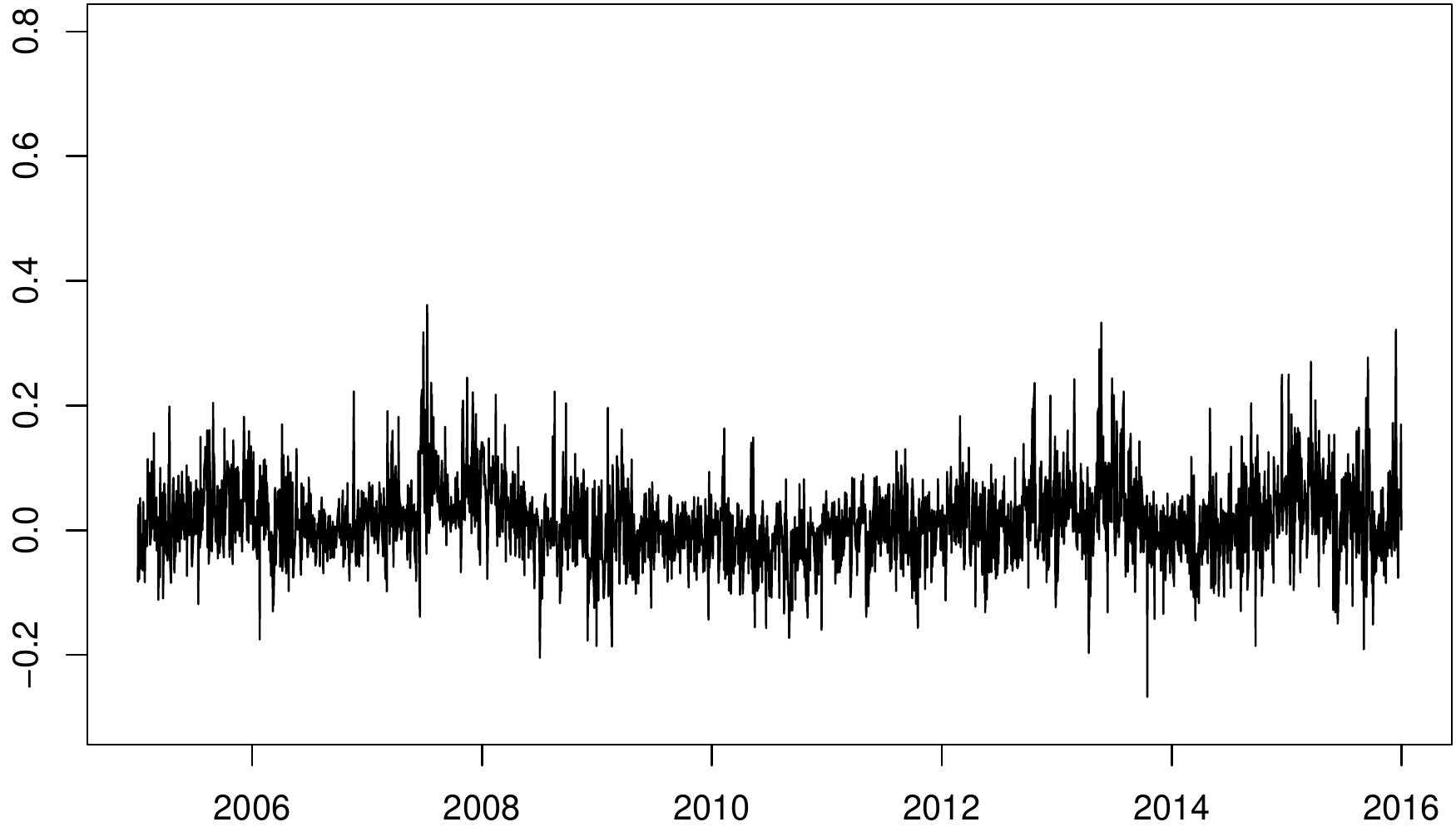}\\

\hspace{0.2cm}\noindent\includegraphics[width=0.48\textwidth, height=0.12\textheight]{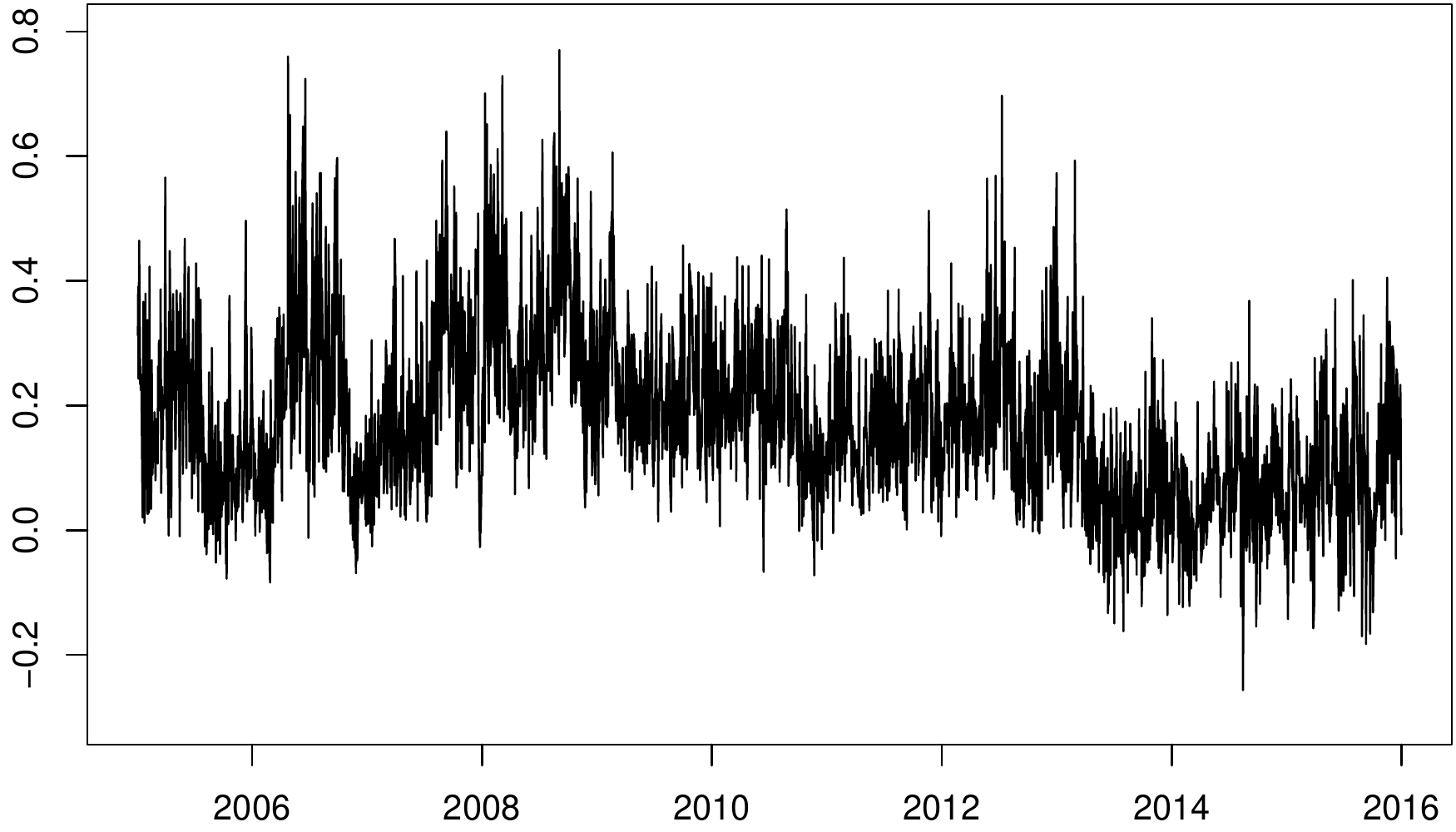}
\noindent\includegraphics[width=0.48\textwidth, height=0.12\textheight]{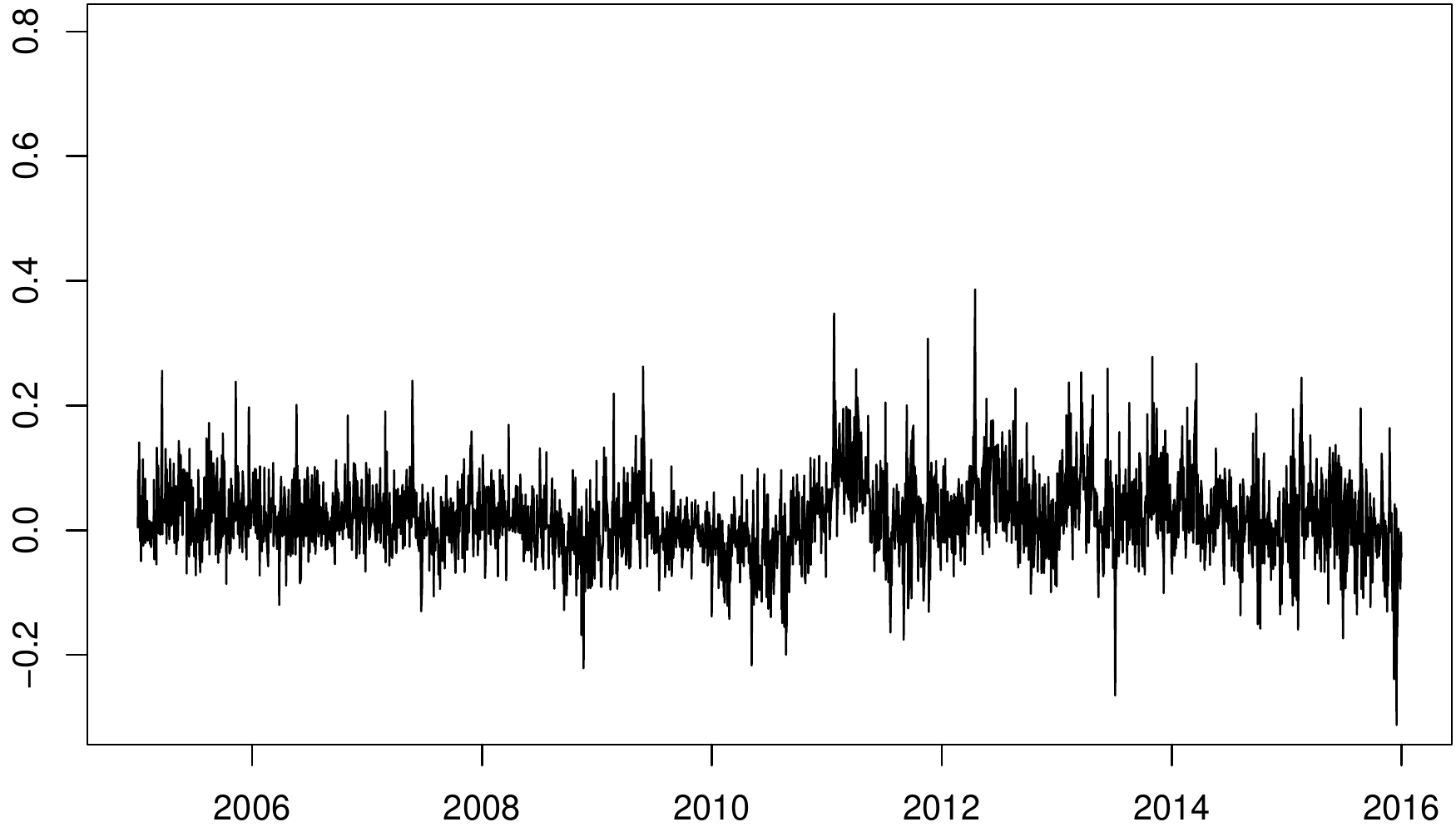}
\vspace{1cm}
\captionsetup{singlelinecheck=off}
\caption[ac]{Graphical representation of cumulative relative importance of sectors over the entire period  2010--2015. One--step ahead weight forecasts for individual stocks in the DJ30 from $\mathsf{DCW}$ are taken in absolute value and then rescaled to sum up to one. Sector values are obtained by aggregation and then ordered (bottom to top) according to the average relative importance; single sector positions are readable as a difference from the lower line (top line =$1$).\medskip}\label{fig:weightstot}
\noindent\includegraphics[width=0.98\textwidth]{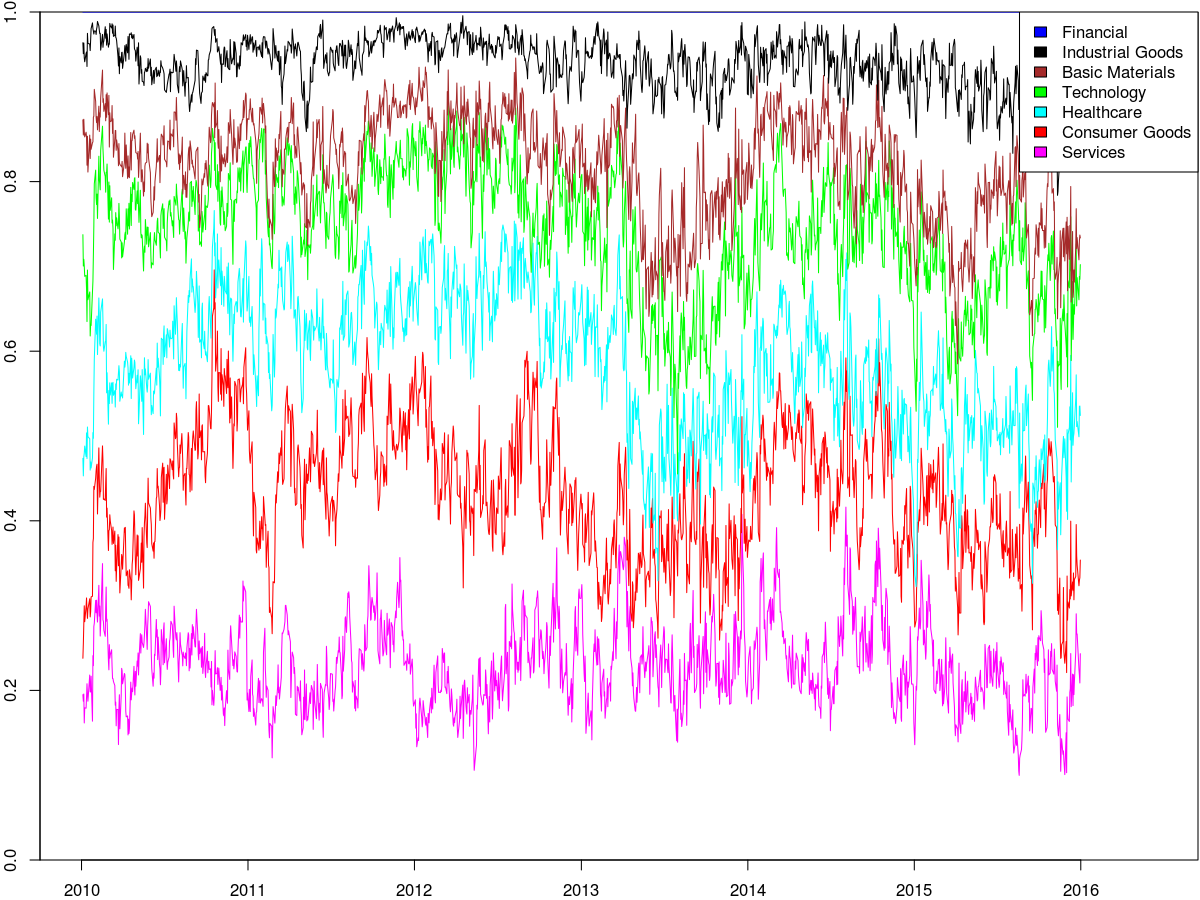}
\end{figure}

Dynamic Conditional Weights ($\mathsf{DCW}$) is an approach directed at the daily time series of realized optimal portfolio weights 
\begin{equation}
\nu_{t}\equiv(\iota'S_{t}^{-1}\iota)^{-1}S_{t}^{-1}\iota.
\label{realwght}
\end{equation} 
The weights $\nu_{t}$ are observable in $t$ (from the observability of the realized $S_{t}$) and minimize the portfolio realized variance $\nu_{t}'S_{t}\nu_{t}$. The time series profile of $\nu_{t}$  for a few tickers may be graphically appraised in Figure \ref{fig:temp} (Apple, Boeing, Johnson and Johnson, and Merck): they display different ranges (same scale is used across) around a changing level, venture into negative territory and, as other financial time series, are characterized by persistence and some short-lived variability.

In order to define the dynamic structure of the $\mathsf{DCW}$ we move from an autoregressive representation of the portfolio-variance minimization problem\footnote{In the Appendix \ref{dcw_general} we derive identical $\mathsf{DCW}$ dynamics from the maximization of a quadratic utility dependent on portfolio returns $r_{t}$. The corresponding realized portfolio weights, which are a function of both returns and var-cov matrix, allow for a seamless merger with the literature focusing on estimation-error reduction in the vector of average returns: see, e.g., the Bayes-Stein shrinkage portfolio of \citet{Jorion:1985} and \citet{Jorion:1986}, the Bayesian portfolio based on belief in an asset-pricing model of \citet{Pastor:2000} and \citet{Pastor:Stambaugh:2000}, the portfolio implied by asset-pricing models with unobservable factors of \citet{MacKinlay:Pastor:2000}, and the three-fund portfolio of \citet{Kan:Zhou:2007}.} and we express it in terms of a time-independent weighting matrix. To see the details, let us recall that $\omega_{t}$ is the vector of weights minimizing the portfolio conditional variance $d\left(\Omega_{t},\omega_{t}\right)\equiv \omega_{t}'\Omega_{t}\omega_{t}$, where $\Omega_{t}=\mathbb{E}_{t-1}\left[S_{t}\right]$ is the time $(t-1)$--conditional expectation of $S_{t}$. Furthermore, let $\left\{d\left(S_{t-i},\omega_{t}\right)\equiv \omega_{t}'S_{t-i}\omega_{t} \right\}_{i=0}^{\infty}$ be the sequence of stationary realized portfolio variances, given the portfolio weights $\omega_{t}$. Thus, the autocorrelation structure of $d\left(S_{t},\omega_{t}\right)$ may be satisfactorily represented as an AR($r$), so that an expression for $d\left(\Omega_{t},\omega_{t}\right)$ can be:
\begin{equation}\label{eq:trun_ar}
d\left(\Omega_{t},\omega_{t}\right) = c_{t} + \sum_{i=1}^{r}\theta_{t,i}d\left(S_{t-i},\omega_{t}\right),
\end{equation}
where $c_{t}$ and $\theta_{t,i}$  (shorthand for $c(\omega_{t})$ and $\theta_{i}(\omega_{t})$ respectively) are such that $c_{t}\ge 0$ and $\theta_{t,i}\ge 0$ $\forall i$ to satisfy necessary and sufficient conditions for the positivity of $d\left(\Omega_{t},\omega_{t}\right)$.

Adding and subtracting $\nu_{t-i}$ to $\omega_t$, the generic element $d\left(S_{t-i},\omega_{t}\right)$ may be rewritten as:
\begin{eqnarray}
d\left(S_{t-i},\omega_{t}\right) & = &\left[\nu_{t-i}+\left(\omega_{t}-\nu_{t-i}\right)\right]'S_{t-i}\left[\nu_{t-i}+\left(\omega_{t}-\nu_{t-i}\right)\right]\nonumber\\
%& \omega_{t}'S_{t-i}\omega_{t} \\
%&&\textrm{adding and subtracting $\nu_{t-i}$ to $\omega_t$, and then substituting} \nonumber \\ 
%& = & \left(\iota'\Omega_{t-i}^{-1}\iota\right)^{-1} + 2\nu_{t-i}'\Omega_{t-i}\left(\omega_{t}-\nu_{t-i}\right)+\left(\omega_{t}-\nu_{t-i}\right)'\Omega_{t-i}\left(\omega_{t}-\nu_{t-i}\right)\\
& = & \left(\iota'S_{t-i}^{-1}\iota\right)^{-1} + 2\left(\iota'S_{t-i}^{-1}\iota\right)^{-1}\iota'\left(\omega_{t}-\nu_{t-i}\right)+\left(\omega_{t}-\nu_{t-i}\right)'S_{t-i}\left(\omega_{t}-\nu_{t-i}\right)\nonumber\\
& = & \left(\iota'S_{t-i}^{-1}\iota\right)^{-1} + \left(\omega_{t}-\nu_{t-i}\right)'S_{t-i}\left(\omega_{t}-\nu_{t-i}\right)\label{eq:sum_q}
\end{eqnarray}
with $\iota'\left(\omega_{t}-\nu_{t-i}\right)=0$ due to the portfolio weights adding to unity by construction. Similarly, in view of the fact that:
\begin{equation*}
1 = \left(\iota'S_{0}^{-1}\iota\right)^{-1} + \left(\omega_{t}-\nu_{0}\right)'S_{0}\left(\omega_{t}-\nu_{0}\right),
\end{equation*}
%the $1$ multiplying the intercept $c$ may be seen as given by $\omega_{t}'\Omega_{0}\omega_{t}$, 
for some symmetric and pd matrix $S_{0}$ and $\nu_{0}=\left(\iota'S_{0}^{-1}\iota\right)^{-1} S_{0}^{-1}\iota$, equation (\ref{eq:trun_ar}) may be rewritten as:
\begin{eqnarray}
d\left(\Omega_{t},\omega_{t}\right) &=& d_{0} + \left(\omega_{t}-\nu_{0}\right)'c_{t} S_{0}\left(\omega_{t}-\nu_{0}\right) + \sum_{i=1}^{r}\left(\omega_{t}-\nu_{t-i}\right)'\theta_{t,i} S_{t-i}\left(\omega_{t}-\nu_{t-i}\right) \nonumber \\
&=& d_{0} + m(\omega_{t})' W_t\, m(\omega_{t}) \label{eq:expand}
\end{eqnarray}
where $d_{0}$ is the sum of all the terms that do not depend on $\omega_{t}$, the vector 
$$
m(\omega_{t})\equiv \left((\omega_{t}-\nu_{0})',(\omega_{t}-\nu_{t-1})',\ldots,(\omega_{t}-\nu_{t-r})'\right)',
$$ 
and the matrix $W_t$ is block-diagonal with blocks $(c_{t}S_{0}, \theta_{t,1}S_{t-1}, \ldots,\theta_{t,r}S_{t-r})$. 

Minimizing $ m(\omega_{t})' W_t\, m(\omega_{t})$ wrt $\omega_t$ in such a context loops back to the usual solution $\omega_{t}=(\iota'\Omega_{t}^{-1}\iota)^{-1}\Omega_{t}^{-1}\iota$, whose implementation requires the separate specification of the dynamics of $\Omega_{t}$.

To avoid modeling $\Omega_{t}$ directly, our suggested solution is to apply the minimization problem to the expression $m(\omega_{t})'Wm(\omega_{t})$ with a time-independent $W$, so that the associated first order conditions  may be written as $Q\,m(\omega_{t})=0$ with $Q=(Q_{0},Q_{1},Q_{2},\ldots,Q_{r})$ where $Q_{i}$ are $(M\times M)$ matrices that do not depend on $\omega_{t}$. Solving for $\omega_{t}$ gives:
\begin{equation*}
\omega_{t} = \widetilde{a} + \sum_{i=1}^{r}\widetilde{A}_{i}\nu_{t-i}
\end{equation*}
where parameters are in the ($M\times 1$) vector $\widetilde{a}$ and in the  ($M\times M$) matrices $\widetilde{A}_{i}$. 

By analogy to other financial time series models, one can replace this AR($r$) representation (presumably needing a large $r$) with a more parsimonious representation for the vector of expected portfolio weights $\omega_{t}$ as an ARMA($p,q$)\footnote{While, in general, a ($p,q$) parameterization does not necessarily coincide with the ($\infty$) parameterization, just like in the case of a GARCH($p,q$) {\it vs} an ARCH($\infty$), the former will provide more stable estimates and accurate forecasts than the latter, all the more when truncated.}:
\begin{equation}\label{eq:dcw_kappa}
\omega_{t} = \kappa + \sum_{i=1}^{p}A_{i}^{*}\nu_{t-i} + \sum_{j=1}^{q}B_{j}^{*}\omega_{t-j}
\end{equation}
which gives the $\mathsf{DCW}$ functional form of the optimal $\omega_{t}$, given the time-independent $W$.  

Finally, taking expectations of both sides of equation (\ref{eq:dcw_kappa}) and letting $\omega\equiv\mathbb{E}[\nu_{t}]$ it follows that:
\begin{equation}\label{eq:dcw_target}
\omega_{t} = \left[I-\sum_{i=1}^{p}A_{i}^{*}-\sum_{j=1}^{q}B_{j}^{*}\right]\omega + \sum_{i=1}^{p}A_{i}^{*}\nu_{t-i} + \sum_{j=1}^{q}B_{j}^{*}\omega_{t-j}
\end{equation}
The specification using full matrix coefficients guarantees $\iota' \omega_{t} = 1$ for all $t$ only if, beside $\iota' \omega = 1$ and $\iota' \omega_{0} = 1$, we impose the restrictions $A^{*\prime}_{i} \iota = a_{i} \iota$ and $B^{*\prime}_{j} \iota = b_{j} \iota$ for all $i$ and $j$ (i.e., in each coefficient matrix all columns must add to the same value). 
This condition is satisfied if the coefficient matrices are scalar but not in the diagonal case, for which the normalization $\omega_{t}/(\iota'\omega_{t})$ is used. Notice that, regardless of the structure of the coefficient matrices, $\mathsf{DCW}$ requires the modeling of only $M$ dynamic components (the portfolio weights) in contrast to standard {\it model-based} approaches which require modeling of $M$ conditional variances and $M(M-1)/2$ conditional correlations.

The parameters in (\ref{eq:dcw_target}) may be estimated using various approaches. Two of them are worth discussing briefly: in the first, estimates are obtained by minimizing the sample portfolio variance itself. This approach has the appealing feature of selecting the model parameters that IS minimize the same function used to evaluate the OOS performance. Its main drawback is that the objective function has to be optimized with respect to all parameters jointly, making the estimation particularly cumbersome and difficult to apply to large cross-sectional dimensions $M$. The second approach is least-squares estimation, performed by the IS minimization of the square distance between predictions $\omega_{t}$ and realizations $\nu_{t}$. When associated to a diagonal parameterization of the matrices $A_{i}^{*}$ and $B_{j}^{*}$, it further allows for the equation-by-equation ARMA estimation of the model parameters. Therefore, in view of its applicability to a vast number of assets $M$, in what follows we focus on the diagonal $\mathsf{DCW}$ specification estimated by least-squares. Furthermore, we will concentrate on the $\mathsf{DCW}$(1,1) parameterization, with a twofold motivation: to provide a fair comparison to the {\it model-based} $\mathsf{DCC}$(1,1) and to work with a baseline specification that is easy to scale to a large number of assets. It follows that the empirical performance of $\mathsf{DCW}$, presented in Section \ref{results}, is likely to improve if a case-by-case best IS specification is derived from any combination of standard selection techniques such as Information Criteria, pruning statistically insignificant parameters, and Box-Jenkins--type procedures based on the properties of the residual ACFs and PACFs. 

\begin{comment}
Indeed, equation (\ref{eq:dcw_kappa}) belongs to the same family of parameterizations as the successful GARCH models and which relate the conditional expectation ($\omega_{t}$) to its lagged values as well as lags of the realizations ($\nu_{t}$) containing the innovations, surprises or errors that drive the conditional expectations. 

\begin{displaymath}
m(\omega_{t})=\left(
\begin{array}{l}
\omega_{t}-\nu_{0}\\
\omega_{t}-\nu_{t-1}\\
\vdots\\
\omega_{t}-\nu_{t-r}\\
\end{array}
\right)
\end{displaymath}

\begin{equation}\label{eq:unc_ar}
d\left(S_{t},\omega_{t}\right) = c_{t} + \sum_{i=1}^{\infty}\theta_{t,i}d\left(S_{t-i},\omega_{t}\right) + u_{t}
\end{equation}

\begin{equation}\label{eq:con_ar}
d\left(\Omega_{t},\omega_{t}\right) = c_{t} + \sum_{i=1}^{\infty}\theta_{t,i}d\left(S_{t-i},\omega_{t}\right)
\end{equation}

\begin{equation*}
Q_{0}(\omega_{t}-\nu_{0}) + \sum_{i=1}^{r}Q_{i}(\omega_{t}-\nu_{t-i}) = 0
\end{equation*}

\begin{equation*}
\omega_{t} = \widetilde{a} + \sum_{i=1}^{\infty}\widetilde{A}_{i}\nu_{t-i}
\end{equation*}
\end{comment}

\section{Empirical Application \label{results}}

The data used for portfolio selection pertain to $M=28$ of the $30$ constituents of the Dow Jones 30 Index. The sample has 11 years of high-frequency daily observations from 01/03/2005 to 12/31/2015 for a total of 2768 days. Two series, with tickers TRV and V, are not included in the study because they are not available for the full sample period\footnote{TRV data are available only from 02/26/2007 while V data are missing from 08/04/2006 to 02/26/2007.}. Tickers of the 28 included stocks are: AAPL, AXP , BA, CAT, CSCO, CVX, DD, DIS, GE, GS, HD, IBM, INTC, JNJ, JPM, KO, MCD, MMM, MRK, MSFT, NKE, PFE, PG, UNH, UTX, VZ, WMT, XOM. The raw tick-by-tick TAQ data are cleaned using the procedure of \citet{Brownlees:Gallo:2006} from which realized kernel covariances are computed following the approach of \citet{BarndorffNielsen:Hansen:Lunde:Shephard:2011}. Details on this procedure may be found in Appendix \ref{data_handling}. The sample is split into six 5-year IS periods: 2005-2009 to 2010-2014, with about 1260 observations each. Each model specification is estimated IS and ensuing OOS forecasts are computed for the following year (about 252 observations).

In Section \ref{PW_stats}, we discuss the portfolio weights forecast performance for the diagonal $\mathsf{DCW}$(1,1) of equation (\ref{eq:dcw_target}), the scalar $\mathsf{DCC}$(1,1) of equation (\ref{eq:dcc}) and the plain $\mathsf{RW}$ of Section \ref{model_free}. $\mathsf{DCW}$ and $\mathsf{DCC}$ with full matrices of coefficients are not considered due to their limited applicability to large cross-sectional dimensions. On the other hand, scalar $\mathsf{DCW}$ and diagonal $\mathsf{DCC}$ are estimated but not reported as their OOS performance is inferior to that of diagonal and scalar, respectively. Instead, the choice of plain $\mathsf{RW}$ as benchmark of the {\it model-free} approaches is motivated by the findings of \citet{Fan:Furger:Xiu:2016}, Figure 5, and \citet{Ait-Sahalia:Xiu:2017}, Figure 6, which show that none of the proposed alternatives consistently outperforms the plain $\Omega_{t}=S_{t-1}$ in attaining the minimum OOS portfolio variance.

In Sections \ref{PV}-\ref{util_env} we comment on the resulting portfolio performances in terms of various standard measures as in \citet{Bollerslev:Patton:Quaedvlieg:2018} and the novel {\it break-even transaction costs}.

\subsection{Portfolio Weights\label{PW_stats}}

Table \ref{tab:dcw_summary_stats} reports descriptive statistics of the equation-by-equation ARMA estimates of the parameters of the diagonal $\mathsf{DCW}$(1,1) of equation (\ref{eq:dcw_target}). Persistence, estimated by $A^{*}+B^{*}$, is in-line with that of realized variances, with $B^{*}$ substantially larger than $A^{*}$. Specifically, over the 168 estimates, the maximum $A^{*}$ is $0.35$ while the minimum $B^{*}$ is approximately $0.5$. Descriptive statistics for the IS $R^{2}$ may be found in Table \ref{tab:dcw_summary_stats_r2}. Over the six IS periods, portfolio weights exhibit different degrees of predictability with $R^{2}$ ranging from $4\%$ to $50\%$. Overall predictability of realized portfolio weights is attested by the average $R^{2}$ which ranges between $20\%$ and $28\%$, depending on the IS period. Given that each realized weight is made of 756 covariances and 28 variances, reported $R^{2}$ are found to be in line with those reported by the {\it model-based} literature for the modeling of realized variances (higher) and covariances (lower).

We measure OOS performance in terms of $R^{2}$ as in \citet{Welch:Goyal:2008}:
\begin{equation*}
R^{2} = 1 -\frac{MSE_{A}}{MSE_{N}}
\end{equation*}
where the $MSE_{A}$ is the OOS mean-squared forecasting-error of the model whose weights forecasts are being evaluated and $MSE_{N}$ is the {\it reference} measure. Notice that, in comparing the performance of competing specifications, the rankings of the OOS $R^{2}$ are unaffected by the choice of $MSE_{N}$. Here, we calculate $MSE_{N}$ with respect to the {\it ex-post} OOS mean of the portfolio weights. Summary statistics of $\mathsf{DCW}$ performance are presented in Table \ref{tab:dcw_summary_stats_r2_out}, with forecasts explaining $10\%$ of the OOS variability in the realized portfolio weights, on average. While for some stocks, $\mathsf{DCW}$ forecasts explain more than $20\%$, for others they are as low as $-7\%$. OOS $R^{2}$ associated to portfolio weights forecasts from $\mathsf{DCW}$, $\mathsf{DCC}$ and $\mathsf{RW}$ are summarized by the kernel density estimates in Figure \ref{fig:all_R2}. It clearly emerges that the OOS weights forecasts of $\mathsf{DCW}$ are superior to those of $\mathsf{DCC}$, which exhibit $R^{2}$ centered at zero and a long left tail. The approximately symmetric distribution of $\mathsf{RW}$ OOS $R^{2}$, centered around $-40\%$, provides further evidence of the relatively poor performance of random walk dynamics.

\begin{figure}[p]
\captionsetup{singlelinecheck=off}
\caption[ac]{
%\scriptsize{
Graphical representation of cumulative relative importance of sectors by year 2010 to 2015. One--step ahead weight forecasts for individual stocks in the DJ30 from $\mathsf{DCW}$ are taken in absolute value and then rescaled to sum up to one.
Sector values are obtained by aggregation and then ordered (bottom to top) according to the average relative importance; single sector positions are readable as a difference from the lower line (top line =$1$).\medskip}
\label{fig:weightsyear}%}
\noindent\includegraphics[width=0.48\textwidth]{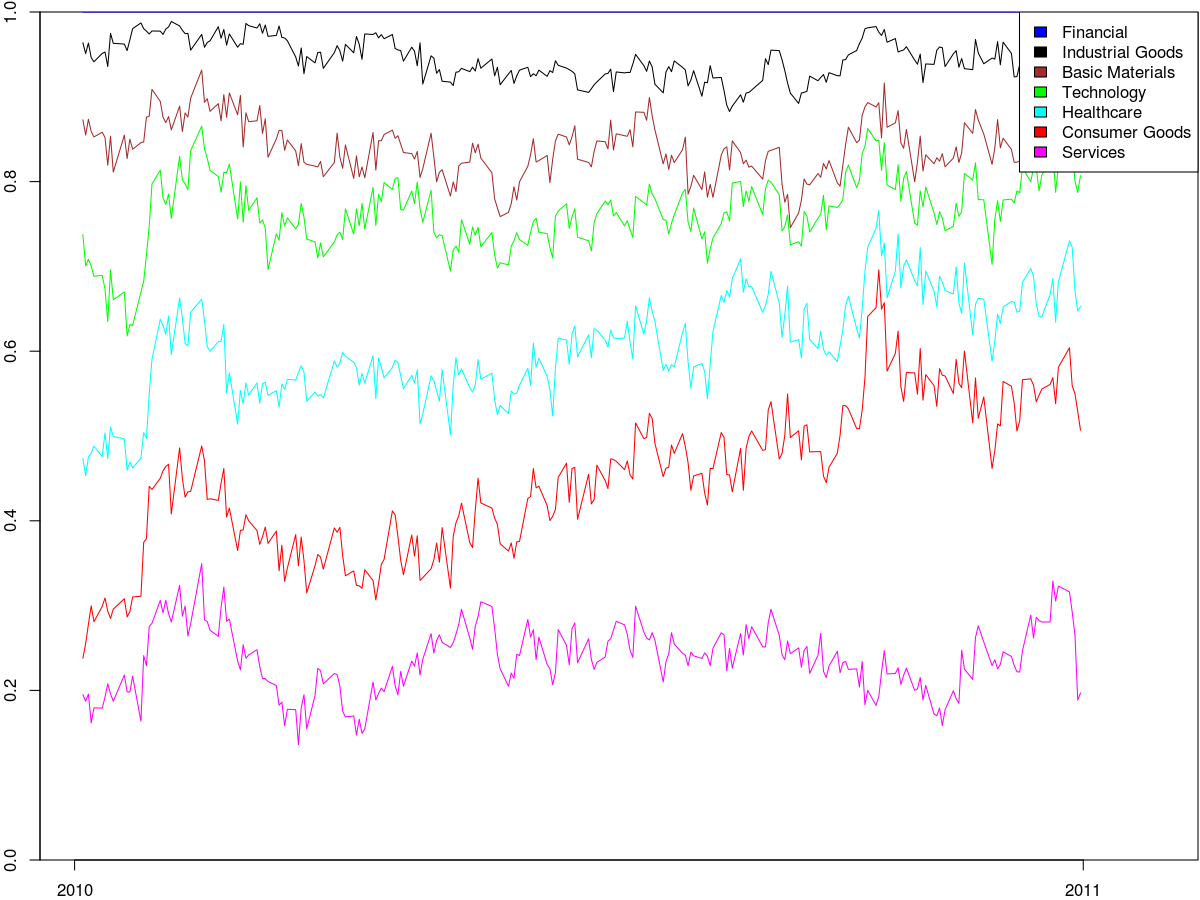}
\hspace{0.5cm}
\includegraphics[width=0.48\textwidth]{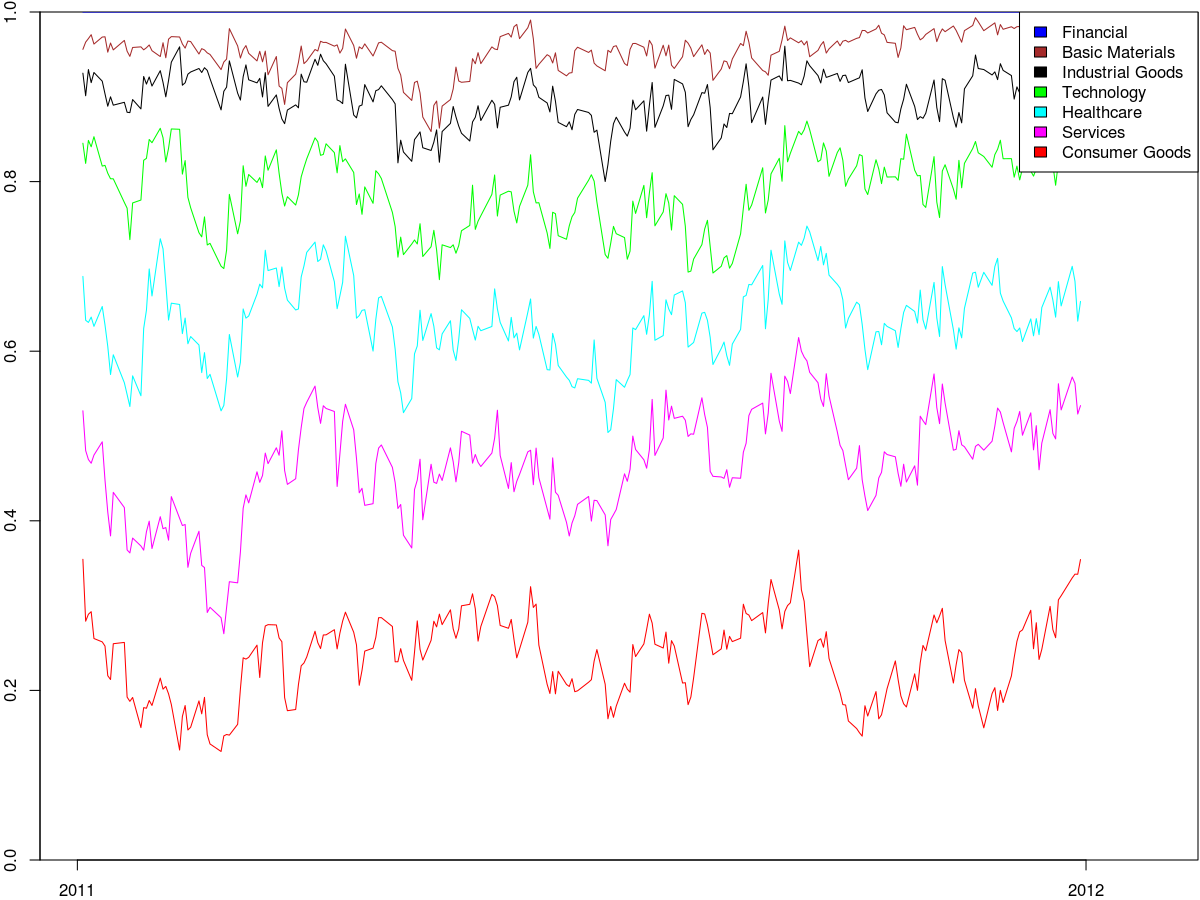}
\vspace{0.5cm}
\includegraphics[width=0.48\textwidth]{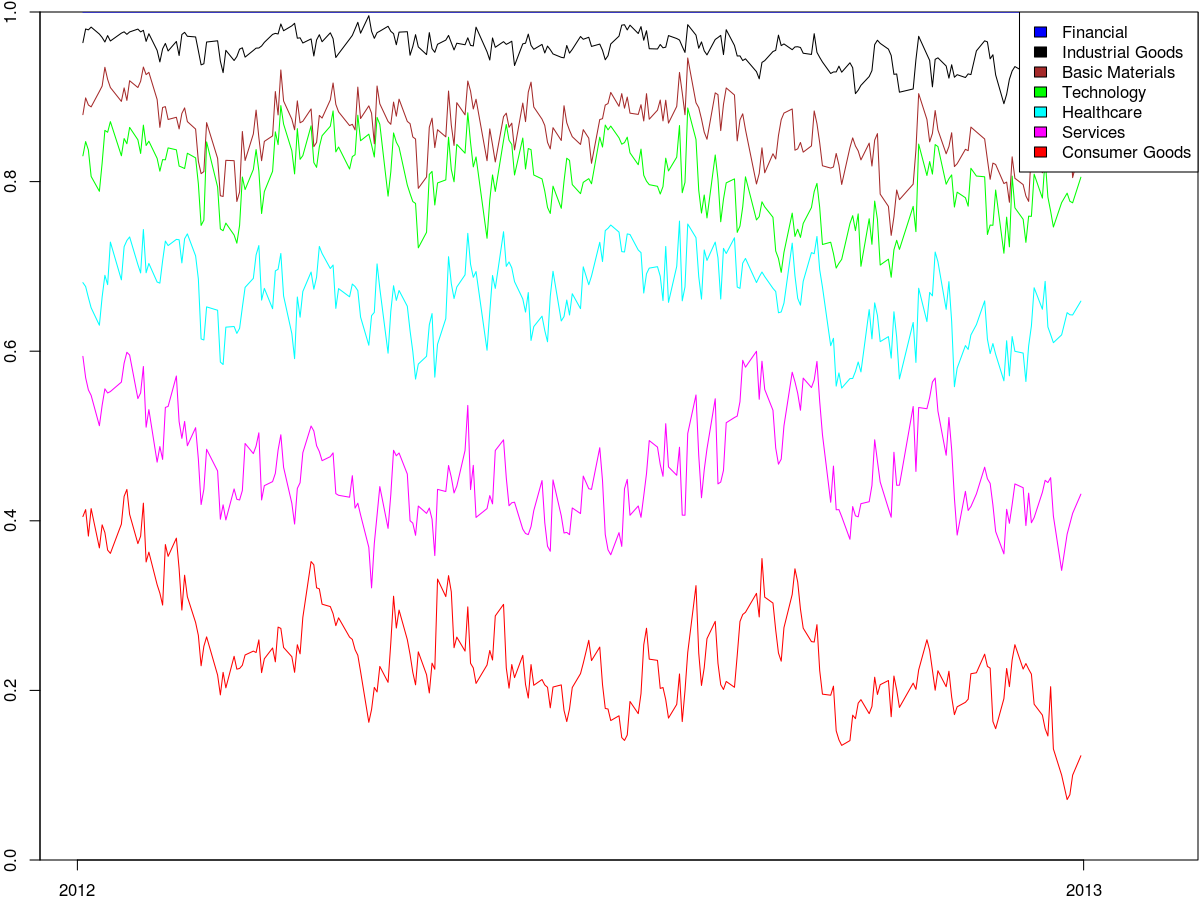}
\hspace{0.5cm}
\includegraphics[width=0.48\textwidth]{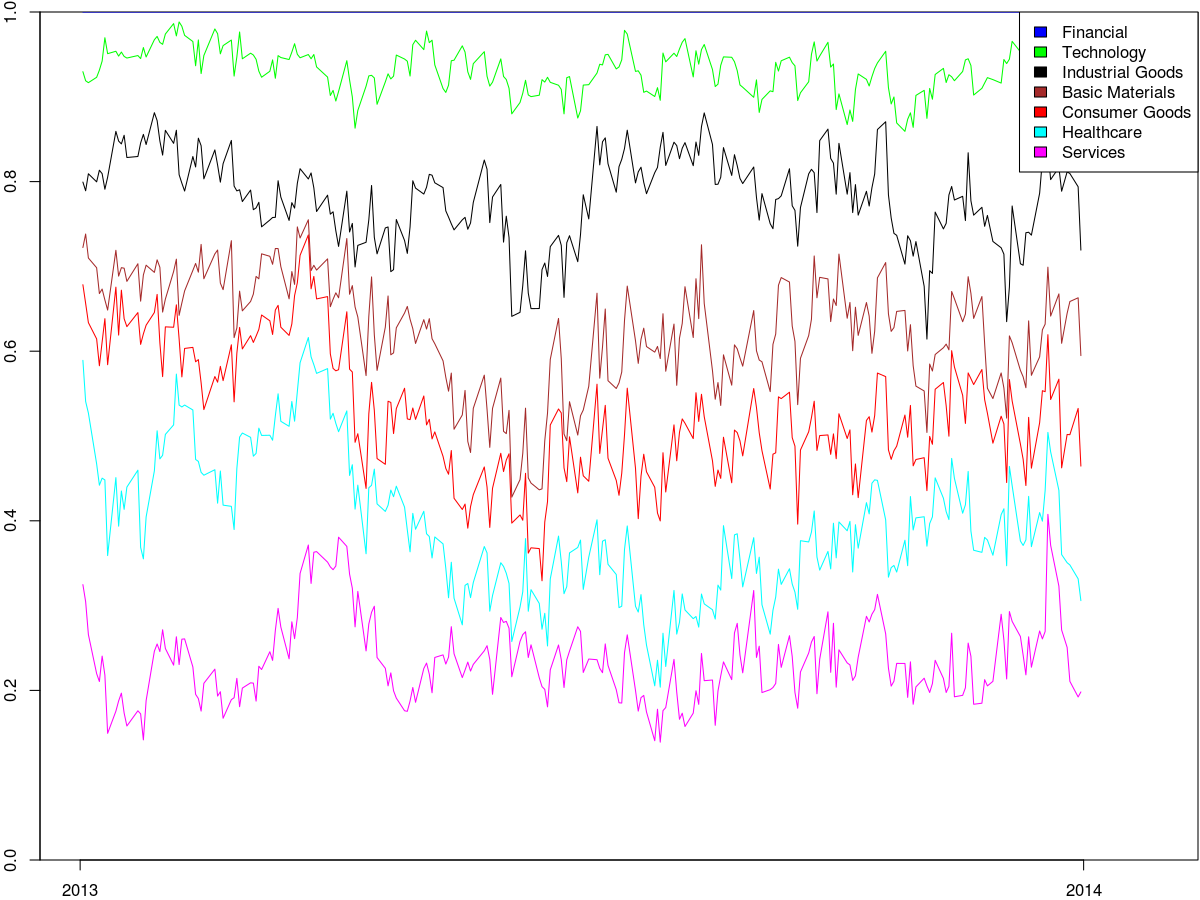}
\vspace{0.5cm}
\includegraphics[width=0.48\textwidth]{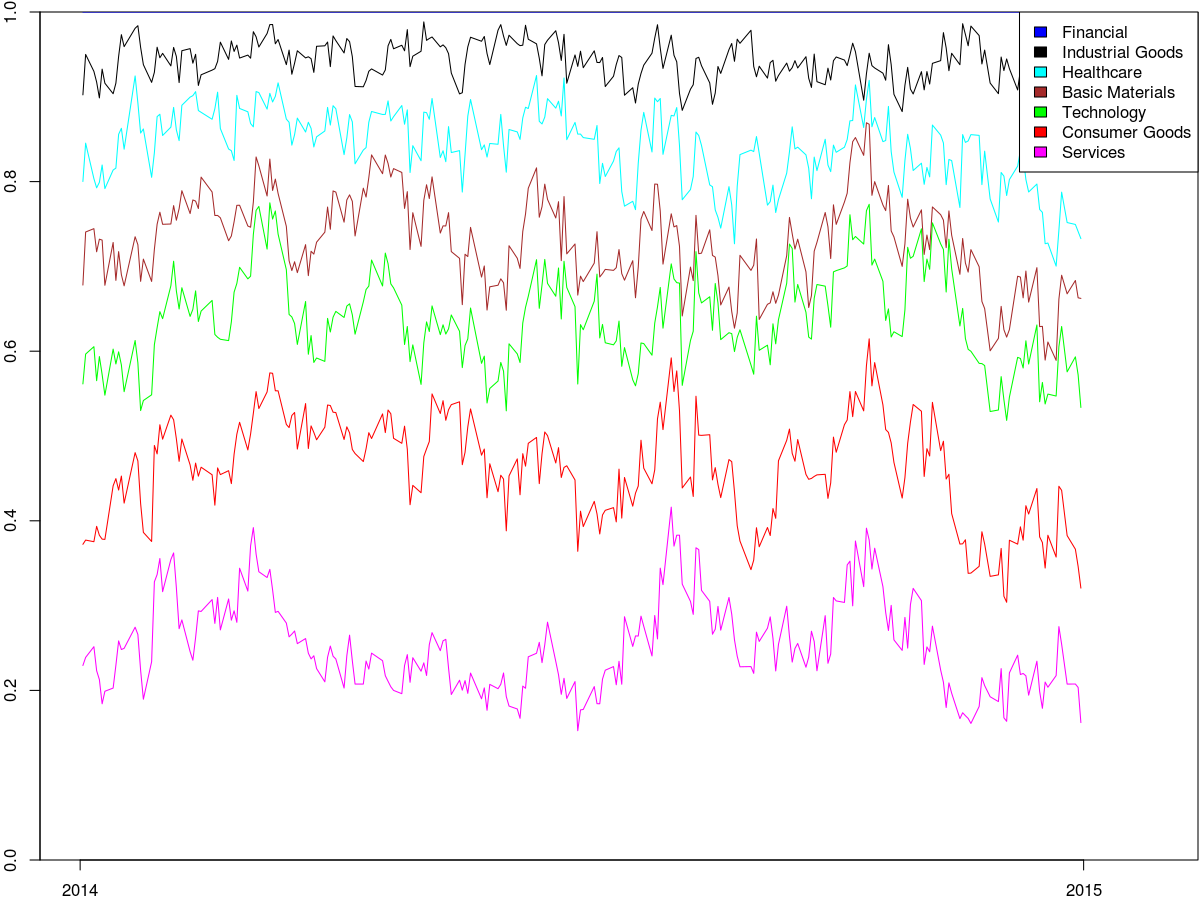}
\hspace{0.5cm}\includegraphics[width=0.48\textwidth]{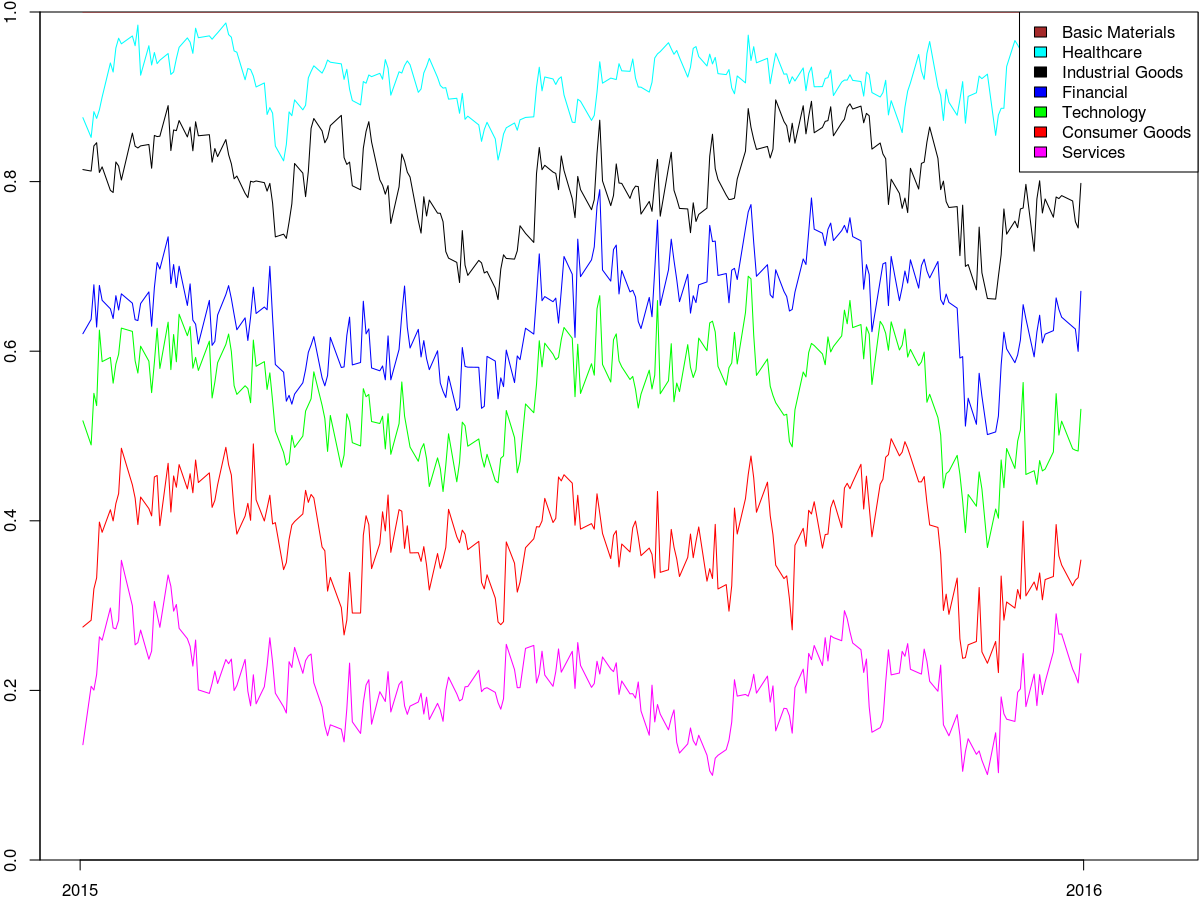}
\vspace{0.5cm}
\end{figure}

To see how the $\mathsf{DCW}$ with $EC=\infty$ forecasts behave in practice, we organize one--step ahead results for individual stocks by taking their absolute value and rescaling them to sum up to one. The outcome is then aggregated by sector and ordered according to the average importance over the period considered. The graphical representation of the cumulative relative importance of sectors (value) is influenced by the corresponding cardinality (i.e. value = average\,$\times$\,\# of tickers); each sector position is readable as the difference from the lower line (the top line being $1$). Over the entire period 2010--2015 (Figure \ref{fig:weightstot}), the relative importance of Services is fairly stable around $0.23$; the next sector is Consumer Goods whose importance oscillates around  $0.20$, although it shows a higher variability and a temporary diminished importance during 2013; Healthcare is next $0.15$ and it shows a diminishing importance with a drastic reduction of its values right after the beginning of 2013. Technology has an average importance of $0.16$ with a fairly stable value over the whole period; Basic Materials has a relative importance of $0.08$; Industrial Goods has an overall value $0.12$: its relative importance seems to increase after the beginning of 2013 for about one year, and then, again, during the first half of 2015. Finally, Financials has a relative importance of $0.06$. Breaking the results by year (Figure \ref{fig:weightsyear}), we get a more detailed view of the evolution of this relative importance: first and foremost the confirmation that Services and Consumer Goods alternate in the top position (4, respectively 2 times). Financial is always in the weakest position (with a substantial gain in 2015); Health Care is fairly prominent in the first four years (reaching the second ranking in 2013), but it rapidly deteriorates in 2014 and even more so in 2015. Technology jumps to the third position in 2014 and 2015.

\subsection{Portfolio Variance ($\mathsf{PV}$) \label{PV}}

One measure of OOS performance is the average portfolio variance\footnote{A common measure of the OOS portfolio performance is the Sharpe ratio which highlights the reward-to-risk. However, given that in this study we concentrate exclusively on the contribution of the conditional second moments to optimal portfolio formation, we deem it more appropriate to use a measure of OOS performance that captures second moment effects only.} that emerges from choosing model $\kappa$:
\begin{equation*}
\mathsf{PV}_{\kappa} = \frac{1}{T}\sum_{t=1}^{T}\widehat{\omega}_{\kappa,t}'S_{t}\widehat{\omega}_{\kappa,t}
\end{equation*}
where $\widehat{\omega}_{\kappa,t}$ is the time $t$ forecast of the optimal portfolio weights from model $\kappa$, $S_{t}$ is the time $t$ realized variance-covariance matrix and $t=1,\ldots,T$ is the OOS period.

From Table \ref{tab:table01}\footnote{From here on, in our comments we focus in particular on the overall results reported in the column labeled \textsc{All}, unless otherwise stated.}, $\mathsf{VT}$ produces smaller portfolio variances than those of the $\mathsf{Naive}$ equally weighted portfolio (by between $7.10\%$ and $17.86\%$, with an average of $14.34\%$ over the six-year period). Overall, the {\it model-free} $\mathsf{RW}$ exhibits $\mathsf{PV}$ improvements between $24.61\%$ ($EC\le\infty$) and $29.04\%$ ($EC\le 1.50$). For any value of $EC$, the portfolio variances of the {\it model-based} $\mathsf{DCC}$ are lower than those of $\mathsf{RW}$ by between $1.53\%$ ($EC\le 1.25$) and $9.08\%$ ($EC\le\infty$). $\mathsf{DCW}$ portfolio variance without exposure constraints ($EC\le\infty$) is the smallest. It is smaller than that of $\mathsf{RW}$ for any value of $EC$, by between $0.80\%$ ($EC\le\infty$) and $13.30\%$ ($EC\le 1.25$). It is smaller than that of $\mathsf{DCC}$ for all $EC$ above $1.50$ (by between $1.16\%$ when $EC\le 1.50$ and $4.65\%$ when $EC\le\infty$), but larger for all $EC$ below $1.25$ (between $0.74\%$ when $EC=1.25$ and $1.53\%$ when $EC=1.00$). Should the investor be able to select the $EC$ parameter {\it ad hoc}, as is the case for some parameters of the {\it model-free} approaches, $\mathsf{RW}$ portfolio variance would be reduced by $3.44\%$ by $\mathsf{DCC}$ and $7.89\%$ by $\mathsf{DCW}$.

\subsection{Certainty Equivalent Return ($\mathsf{CEQ}$) \label{CEQ}}

Another common measure of OOS performance is the certainty equivalent return. It is defined as the certain return that an investor is willing to accept to switch from model $\kappa_{1}$ to $\kappa_{2}$:
\begin{equation}\label{eq:ceq02}
\mathsf{CEQ}_{\kappa_{1}\rightarrow\kappa_{2}} = \gamma\cdot\frac{1}{2}\left(\mathsf{PV}_{\kappa_{1}}-\mathsf{PV}_{\kappa_{2}}\right)
\end{equation}
Reporting $\mathsf{CEQ}_{\kappa_{1}\rightarrow\kappa_{2}}$ for $\gamma=1$ allows for the immediate calculation of the certainty equivalent return for any value of risk-aversion\footnote{For example, De Miguel {\it el al}. (2009) consider risk-aversion coefficients of $\gamma=\{1,2,3,4,5,10\}$.} simply by rescaling the reported value by $\gamma$. While the rankings it generates within this framework are no different from those of $\mathsf{PV}$, $\mathsf{CEQ}$ may still be helpful in quantifying the differences in portfolio variances by translating them into returns.\\

As shown in Table \ref{tab:table02}, $\mathsf{VT}$ exhibits OOS certainty equivalences, with respect to $\mathsf{Naive}$, that range between 0.95 and 6.84 average daily basis points and 3.63 basis points over the whole OOS period. Switching from $\mathsf{VT}$ to $\mathsf{RW}$ the certainty equivalence ranges from 5.33 ($EC\le\infty$) to 6.29 ($EC\le 1.50$) basis points. The switch from $\mathsf{RW}$ to $\mathsf{DCC}$ also exhibits positive $\mathsf{CEQ}$ for any exposure constraint and between 0.24 ($EC\le 1.25$) and 1.48 ($EC\le\infty$) basis points. The switch from $\mathsf{DCC}$ to $\mathsf{DCW}$ exhibits positive $\mathsf{CEQ}$ for $EC$ at and above $1.50$, between 0.17 ($EC\le 1.50$) and 0.69 ($EC\le\infty$) basis points, but -0.11 and -0.24 basis points for ($EC\le 1.25$) and ($EC=1.00$), respectively. Again, should the investor be able to select the $EC$ parameter {\it ad hoc}, switching from $\mathsf{RW}$ to $\mathsf{DCC}$ and $\mathsf{DCW}$ would correspond to 1.48 and 2.17 daily basis points, respectively.

\subsection{Turnover ($\mathsf{TO}$) \label{TO}}

In this study, where the focus is on daily trading with no overnight holdings, we have zero portfolio weights prior to rebalancing. Hence, average turnover is given by
\begin{equation*}
\mathsf{TO}_{\kappa} = \frac{1}{T}\sum_{t=1}^{T}\sum_{j=1}^{M}\left|\widehat{\omega}_{\kappa,j,t}\right|
\end{equation*}
and captures the average portfolio exposure $\overline{EC}_{\kappa}$ of forecasting model $\kappa$: in the optimal portfolio allocation literature it is commonly reported, as it is of relevant interest for investors. This measure, which does not include assets' returns, is derived in Appendix \ref{TOA} where we also show its precision up to two orders of magnitude.

In Table \ref{tab:table03}, $\mathsf{Naive}$ and $\mathsf{VT}$ have turnovers of 1.00, by construction. Of the other strategies, for any exposure constraint $EC$, $\mathsf{RW}$ exhibits the highest turnover\footnote{Turnover is a reflection of the dispersion of the resulting portfolio weights and, in the presence of estimation and forecasting errors, it may be taken as an indirect measure of their magnitude.} $\mathsf{TO}$, followed by $\mathsf{DCC}$ and, last, $\mathsf{DCW}$. In the presence of transaction costs, the implications for the investor are that $\mathsf{RW}$ gives rise to the most expensive allocations, followed by those of $\mathsf{DCC}$ and $\mathsf{DCW}$.

\subsection{Break-Even Transaction Costs ($\mathsf{BETC}$)  \label{BETC}}

To provide a comprehensive view of portfolio performance which includes both $\mathsf{PV}$ and the associated transaction costs from $\mathsf{TO}$, we introduce break-even transaction costs as a novel measure of portfolio performance. Specifically, $\mathsf{BETC}$ identifies the transaction costs for which two portfolio allocations are indifferent and consecutively the range over which one allocation is preferred to the other. As shown in Appendix \ref{TCA}, with markup transaction costs $\tau$, average transaction costs $\overline{TC}_{\kappa}$ for model $\kappa$ may be approximated up to two orders of magnitude by:
\begin{equation*}
\overline{TC}_{\kappa} \approx 2\tau\cdot\mathsf{TO}_{\kappa}
\end{equation*}
which, combined with equation (\ref{eq:ceq02}), allows to derive the {\it net} certainty equivalent return:
\begin{equation}\label{eq:nceq01}
NCEQ_{\kappa_{1}\rightarrow\kappa_{2}} = \gamma\cdot\frac{1}{2}\left(\mathsf{PV}_{\kappa_{1}}-\mathsf{PV}_{\kappa_{2}}\right)+2\tau\left(\mathsf{TO}_{\kappa_{1}}-\mathsf{TO}_{\kappa_{2}}\right)
\end{equation}
The {\it break-even transaction cost} ($\mathsf{BETC}$) is defined as the value of $\tau/\gamma>0$ that sets equation (\ref{eq:nceq01}) to zero:
\begin{equation*}
\mathsf{BETC}_{\kappa_{1}\rightarrow\kappa_{2}} = -\frac{1}{4}\cdot\frac{\mathsf{PV}_{\kappa_{1}}-\mathsf{PV}_{\kappa_{2}}}{\mathsf{TO}_{\kappa_{1}}-\mathsf{TO}_{\kappa_{2}}},
\end{equation*}
and hence it combines  $\mathsf{PV}$ and $\mathsf{TO}$ to identify transaction costs per units of risk-aversion ($\tau/\gamma$) for which one approach is preferred to another. $\mathsf{BETC}$ are reported in Table \ref{tab:table04}: entries may be simply multiplied by $\gamma$ to obtain transaction costs corresponding to risk-aversions different from unity. $\mathsf{VT}$ is preferred to $\mathsf{Naive}$ for any level of the transaction costs $\tau$: smaller $\mathsf{PV}$ and equal $\mathsf{TO}$. $\mathsf{RW}$ is preferred to $\mathsf{VT}$ for any $\tau$ only with no short-selling constraints $EC=1.00$. In the other cases, $\mathsf{RW}$ is preferred to $\mathsf{VT}$ for greater risk-aversion $\gamma$ and non-negligible transaction costs. Both $\mathsf{DCC}$ and $\mathsf{DCW}$ are preferred to $\mathsf{RW}$ for any $\gamma$ and $\tau$. This is due to the fact that their (estimated) shrinkage produces smaller $\mathsf{PV}$ and lower $\mathsf{TO}$, both indicative of portfolio weights of higher quality. $\mathsf{DCW}$ is always preferred (any $\tau/\gamma$) to $\mathsf{DCC}$ for $EC$ at and above $1.50$. It is interesting to note that the year 2010 contains some influential data connected to the flash crash of May 6 which impact on the results: while we opted for not arbitrarily correcting for those specific values, the overall  results on 2011--2015 confirm that $\mathsf{DCW}$ is to be preferred\footnote{Excluding 2010, for the triplet $\mathsf{RW}$, $\mathsf{DCC}$ and $\mathsf{DCW}$ we have $\mathsf{PV}=\{0.293206, 0.289647, 0.287181\}$, for $EC\le 1.25$, and $\mathsf{PV}=\{0.307414, 0.301346, 0.300040\}$, for $EC=1.00$. Similarly, $\mathsf{TO}=\{1.23, 1.23, 1.09\}$, for $EC\le 1.25$, and $\mathsf{TO}=1$ for all three when $EC=1.00$. Hence, there is no $\tau/\gamma$ for which $\mathsf{RW}$ is preferred to $\mathsf{DCC}$ or $\mathsf{DCC}$ is preferred to $\mathsf{DCW}$.} to $\mathsf{RW}$ and $\mathsf{DCC}$ for any $EC$ and $\tau/\gamma$.

\subsection{Utility Levels \label{util_env}}
\begin{figure}	
\vspace{1cm}
\captionsetup{singlelinecheck=off}
\caption[ac]{
Density representation of OOS $R^{2}$ of forecasted portfolio weights across assets and time periods.\medskip}\label{fig:all_R2}
\noindent\includegraphics[width=\textwidth, height=0.35\textheight]{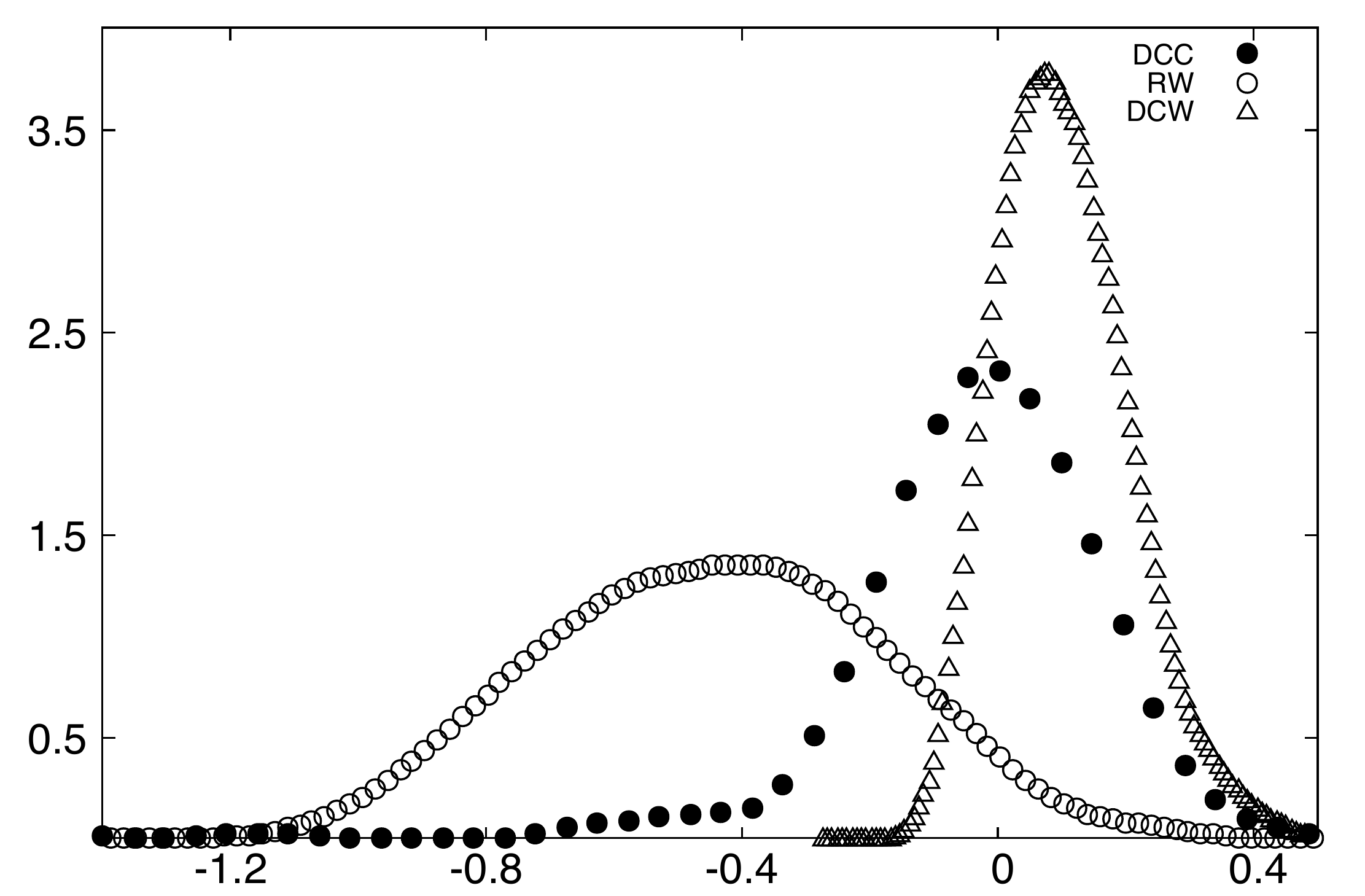}
\vspace{1cm}
\captionsetup{singlelinecheck=off}
\caption[ac]{
%\scriptsize{
Graphical representation of utility envelopes of $\mathsf{RW}$, $\mathsf{DCC}$ and $\mathsf{DCW}$ with respect to $\mathsf{VT}$. On the horizontal axis are the transaction costs per units of risk-aversion $\tau/\gamma$ and on the vertical axis are the utilities measured with respect to that of $\mathsf{VT}$. The first graph represents the entire period 2010--2015, while the second graph excludes the year 2010.\medskip}\label{fig:delta_utility}
\noindent\includegraphics[width=0.49\textwidth, height=0.25\textheight]{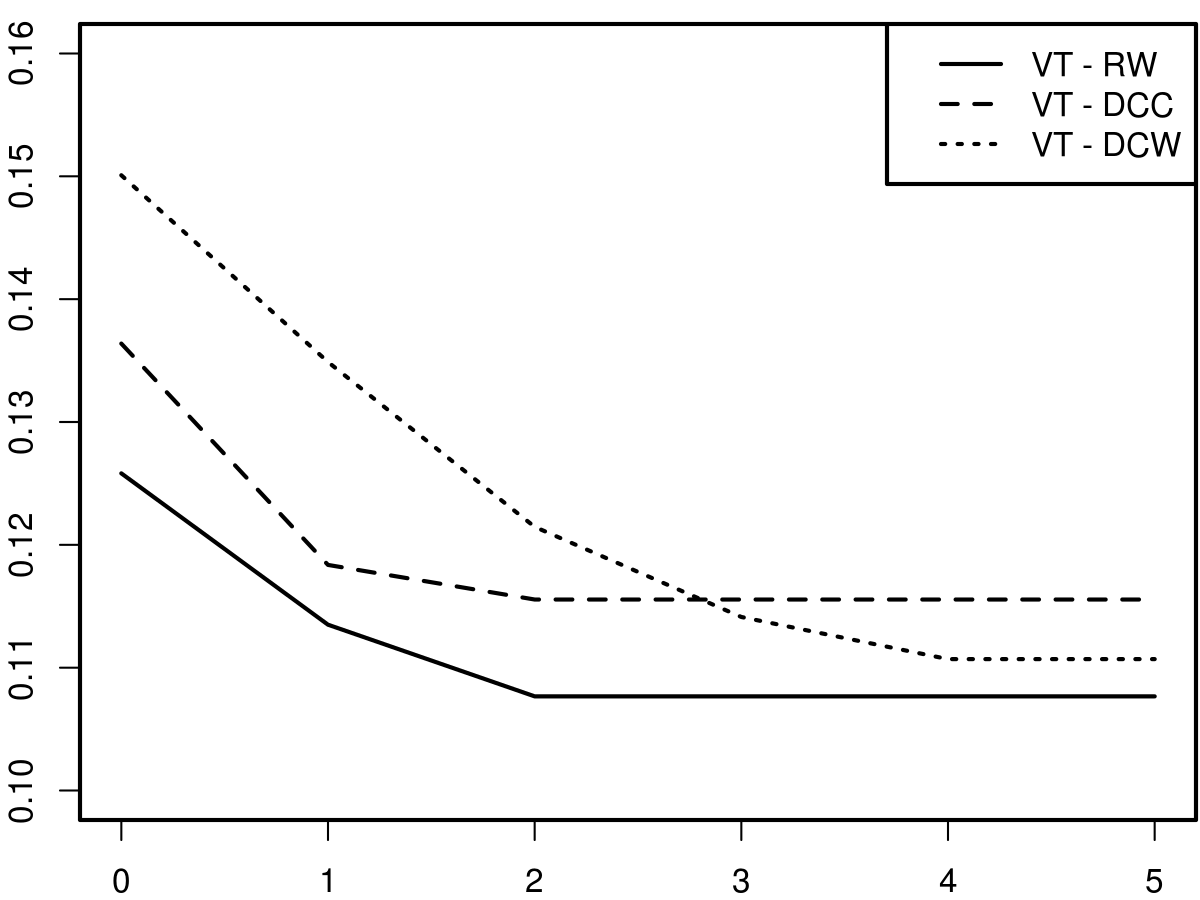}
\noindent\includegraphics[width=0.49\textwidth, height=0.25\textheight]{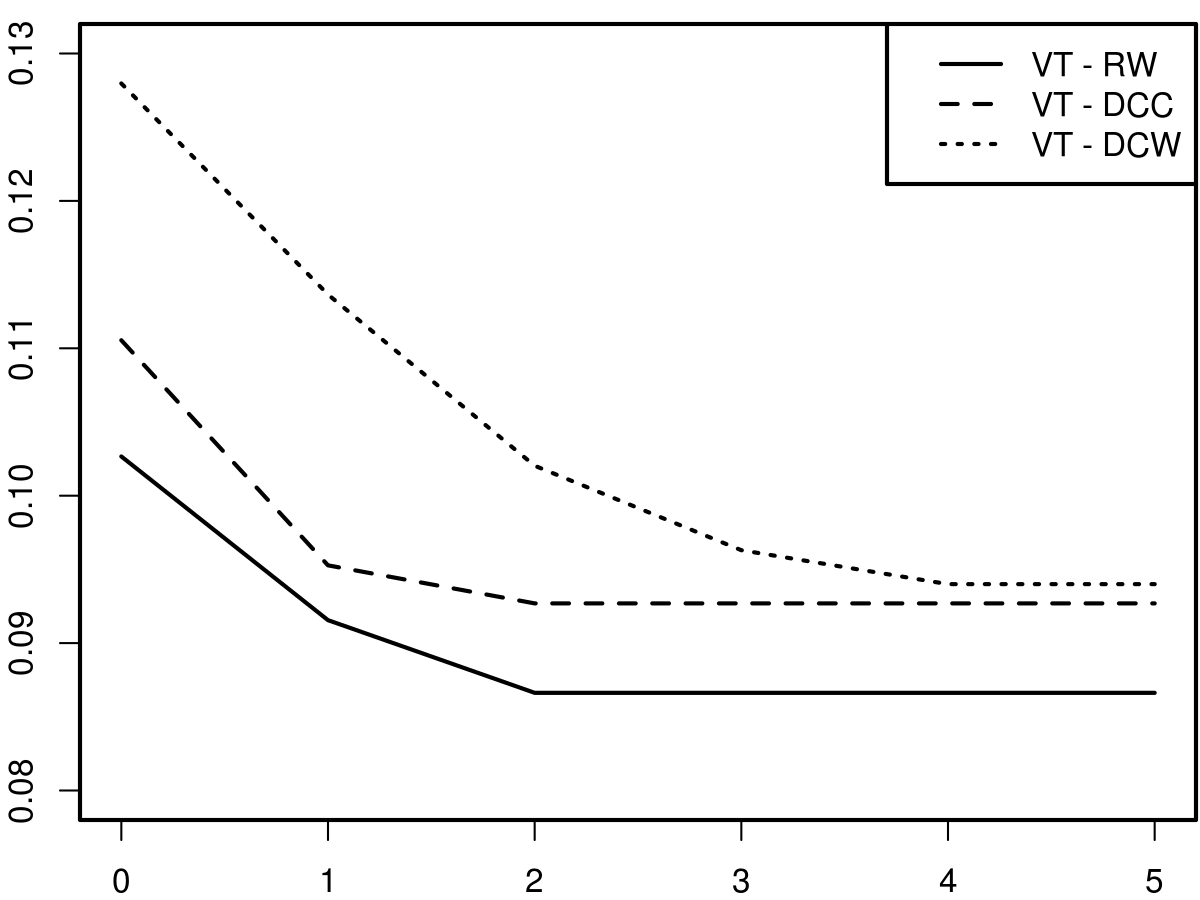}
\end{figure}

In Figure \ref{fig:delta_utility} we report the utility levels\footnote{In the presence of transaction costs, the utility function of equation (\ref{eq:portfolio_utility}) becomes $V_{t} = -2\tau\mathsf{TO}_{\kappa}-0.5\gamma\mathsf{PV}_{\kappa}$. Reported utility levels are those associated to the {\it rank-preserving} transformation $\widetilde{V}_{t} = -2(\tau/\gamma)\mathsf{TO}_{\kappa}-0.5\mathsf{PV}_{\kappa}$.} associated with the various approaches. Specifically, for a given strategy, we begin by plotting the utilities associated with a given strategy as a function of $\tau/\gamma$ for the various exposure constraints $EC$. We then construct the envelope of each strategy as a function of $\tau/\gamma$. The envelopes are piecewise linear curves (the utilities are linear in $\tau/\gamma$) which identify the maximal utility attainable by each strategy for the {\it ex post} optimal exposure constraint $EC$. Finally, we report the envelope differences of $\mathsf{RW}$, $\mathsf{DCC}$ and $\mathsf{DCW}$ with respect to $\mathsf{VT}$. For low $\tau/\gamma$, high $EC$ allocations are preferred as they produce portfolios with smaller variances. On the other hand, when $\tau/\gamma$ is high, low $EC$ allocations are preferred as the increase in portfolio variance is more than compensated by the decrease in the associated transaction costs. From the first graph of Figure \ref{fig:delta_utility}, $\mathsf{DCC}$ and $\mathsf{DCW}$ outperform $\mathsf{RW}$ for any $\tau/\gamma$, while $\mathsf{DCC}$ is preferred to $\mathsf{DCW}$ for $\tau/\gamma$ greater than $3$ basis points. Once again, excluding the year $2010$ produces slightly different results as shown in the bottom panel of Figure \ref{fig:delta_utility}: $\mathsf{DCW}$ is preferred to both $\mathsf{DCC}$ and $\mathsf{RW}$, for any $\tau/\gamma$. Furthermore, as $\tau/\gamma$ increases and $EC=1.00$ becomes optimal for all approaches, the allocation gains of $\mathsf{DCW}$ over $\mathsf{DCC}$ tend to vanish while their difference with respect to $\mathsf{RW}$ remains sizeable. 

\begin{comment}
\subsection{$\mathsf{DCW}$ Portfolio Composition}
\end{comment}

\section{Conclusions \label{concl}}

In this paper we motivate the use of Dynamic Conditional Weights by deriving its dynamic structure by expressing the autoregressive representation of the portfolio-variance minimization problem in terms of a time-independent weighting matrix. We evaluate portfolio weights forecasts from the proposed approach against those of representative {\it model-based} and {\it model-free} specifications which forecast conditional var-cov matrices to calculate the optimal weights. Specifically, the scalar $\mathsf{DCC}$ with HAR variance dynamics as a manageable representative model for the {\it model-based} class and the daily $\mathsf{RW}$ as the benchmark specification for the {\it model-free} class. We find the $\mathsf{DCW}$ portfolio allocations to have lower variance $\mathsf{PV}$ and turnover $\mathsf{TO}$ than $\mathsf{DCC}$ and $\mathsf{RW}$, for any value of the exposure constraints $EC$. The proposed $\mathsf{BETC}$ criterion, which allows the joint evaluation of strategy performance and implementation costs, highlights that, for any realistic level of transaction costs, investors would switch from $\mathsf{RW}$ to $\mathsf{DCC}$  and, with the exception of the no short-selling case $EC=1.00$, would switch from $\mathsf{DCC}$ to $\mathsf{DCW}$.

While measures of portfolio performance are of primary interest in a portfolio management framework, our analysis suggests not to overlook the forecasts of the portfolio weights. In fact, considering that portfolio measures not only capture how close the weights forecasts are to the {\it realizations}, but also their diversification effects, a given strategy may perform relatively well because it provides one of the infinitely many good diversifications despite poorly forecasting the optimal weights. This may be the case for the $\mathsf{RW}$ at the heart of the {\it model-free} approaches: considering the lack of evidence supporting random walk dynamics for variances and covariances and the relatively poor performance of the associated weights forecasts, performance of the $\mathsf{RW}$ portfolio allocations may mostly reflect diversification.

The Dynamic Conditional Weights approach is readily extendible (Appendix \ref{dcw_general}) to the general case of a quadratic utility maximization to incorporate the advances in the reduction of estimation-error in the vector of average returns as in \citet{DeMiguel:Martin-Utrera:Nogales:2014}, \citet{Bouaddi:Taamouti:2013}, \citet{Behr:Guettler:Miebs:2013}, \citet{Behr:Guettler:Truebenbach:2012}, \citet{Kirby:Ostdiek:2012}, \citet{Tu:Zhou:2011}, \citet{DeMiguel:Plyakha:Uppal:Vikov:2010} and \citet{Brandt:Santa-Clara:Valkanov:2009}, among others. Another noteworthy extension of the proposed approach is the one that ensues when the weighting matrix is allowed to exhibit {\it time-dependence}. The resulting $\mathsf{DCW}$ dynamics will display a higher degree of flexibility by allowing for time-varying parameters, a route suggested by \citet{Bollerslev:Patton:Quaedvlieg:2016} to alleviate model misspecification in the context of var-cov modeling. We leave these and other refinements to future research.

\begin{comment}
Hybrid combinations of different features of the various approaches may provide further improvements in this context: specifically, any shrinkage from the {\it model-free} literature may be substituted for the plain realizations of the {\it model-based} approaches or, equivalently, the dynamics from any of the {\it model-based} approaches may be substituted for the random walk assumption of the {\it model-free} approaches. Similar combinations may be envisaged for $\mathsf{DCW}$ and the {\it model-free} approaches. In particular, to apply $\mathsf{DCW}$ when the realizations are not pd, one could choose from an array of possibilities including, but not limited to, an exponential smoothing of the realizations and any of the shrinkage approaches of the {\it model-free} literature. 
\end{comment}

% \bibliographystyle{apecon}
% \bibliography{biblio}

\newpage
\appendix\section{Appendix \label{appdx_A}}

\newtheorem{theorem}{Theorem}[section]
\newtheorem{lemma}{Lemma}[section]
\newtheorem{corollary}{Corollary}[section]
\newtheorem{definition}{Definition}[section]

\subsection{Dynamic Conditional Weights for Quadratic Utility\label{dcw_general}}

Consider the problem of an investor forecasting portfolio weights $\omega_{t}$ in $(t-1)$ to maximize the quadratic utility:
\begin{equation*}
V(r_{t},S_{t},\omega_{t}) = \omega_{t}'r_{t}-\frac{\gamma}{2}\omega_{t}'S_{t}\omega_{t} 
\end{equation*}
where $r_{t}$ and $S_{t}$ are the realized vector of returns and var-cov matrix, respectively. Let $\left\{V(r_{t-i},S_{t-i},\omega_{t})\right\}_{i=0}^{\infty}$ be the time series of realized utilities $V(r_{t-i},S_{t-i},\omega_{t})=\omega_{t}'r_{t-i}-0.5\gamma\omega_{t}S_{t-i}\omega_{t}$, reconstructed at time $t$, given the portfolio weights $\omega_{t}$. Following the same steps of Section \ref{weights}, assume the autocorrelation structure of $V(r_{t},S_{t},\omega_{t})$ is captured, with the desired degree of precision, by an AR($r$) and take $\mathbb{E}_{t-1}$ of both sides of the equality:
\begin{equation}\label{eq:dcw_appdx_01}
V\left(\mathbb{E}_{t-1}(r_{t}),\Omega_{t},\omega_{t}\right) = c_{t} + \sum_{i=1}^{r}\theta_{t,i}V\left(r_{t-i},S_{t-i},\omega_{t}\right)
\end{equation}
where $c_{t}\ge 0$ and $\theta_{t,i}\ge 0$, $\forall i$ to guarantee positivity of the conditional utility $V\left(\mathbb{E}_{t-1}(r_{t}),\Omega_{t},\omega_{t}\right)$. Let $\delta_{t,i}=(\omega_{t}-\nu_{t-i})$ where $\nu_{t-i}=\gamma^{-1}S_{t-i}^{-1}r_{t-i}$ is the vector of realized optimal portfolio weights that maximize $V(r_{t-i},S_{t-i},\nu_{t-i})$, then the generic element $V\left(r_{t-i},S_{t-i},\omega_{t}\right)$ may be rewritten as:
\begin{eqnarray*}
V\left(r_{t-i},S_{t-i},\omega_{t}\right) & = & \omega_{t}'r_{t-i}-\frac{\gamma}{2}\nu_{t-i}'S_{t-i}\nu_{t-i} - \gamma\delta_{t,i}'S_{t-i}\nu_{t-i} - \frac{\gamma}{2}\delta_{t-i}'S_{t-i}\delta_{t-i}\\
& = & \omega_{t}'r_{t-i}-\frac{\gamma}{2}\nu_{t-i}'S_{t-i}\nu_{t-i} - \delta_{t,i}'r_{t-i} - \frac{\gamma}{2}\delta_{t-i}'S_{t-i}\delta_{t-i}\\
& = & \nu_{t-i}'r_{t-i}-\frac{\gamma}{2}\nu_{t-i}'S_{t-i}\nu_{t-i} - \frac{\gamma}{2}\left(\omega_{t}-\nu_{t-i}\right)'S_{t-i}\left(\omega_{t}-\nu_{t-i}\right)
\end{eqnarray*}
Similarly, in view of the fact that:
\begin{equation*}
1 = \frac{1}{2\gamma}r_{0}'S_{0}^{-1}r_{0}-\frac{\gamma}{2}\left(\omega_{t}-\nu_{0}\right)'S_{0}\left(\omega_{t}-\nu_{0}\right)
\end{equation*}
for some vector $r_{0}$ and some symmetric and pd matrix $S_{0}$ with $\nu_{0}=\gamma^{-1}S_{0}^{-1}r_{0}$, equation (\ref{eq:dcw_appdx_01}) may be rewritten as:
\begin{eqnarray*}
V\left(\mathbb{E}_{t-1}(r_{t}),\Omega_{t},\omega_{t}\right) & = & d_{0} - \frac{\gamma}{2}c_{t}\left(\omega_{t}-\nu_{0}\right)'S_{0}\left(\omega_{t}-\nu_{0}\right)\\
& - & \frac{\gamma}{2} \sum_{i=1}^{r}\theta_{t,i}\left(\omega_{t}-\nu_{t-i}\right)'S_{t-i}\left(\omega_{t}-\nu_{t-i}\right)
\end{eqnarray*}
where $d_{0}$ is the sum of all the terms that do not depend on $\omega_{t}$. Following {\it mutatis mutandis} the steps of Section \ref{weights} leads to the same dynamic specification of $\omega_{t}$ as in equation (\ref{eq:dcw_target}) with the only difference being in the definition of the realized weights $\nu_{t-i}$.

\subsection{Data Handling\label{data_handling}}
For each trading day $t$, let $\{r_{j}\}_{j=1}^{J}$ be the collection of the $(M\times 1)$ return-vectors resulting from price-vectors synchronized according to Barndorff-Nielsen {\it et al.} (2011). The daily realized kernel variance-covariance matrix is then computed as:
\begin{equation*}
S = \sum_{h = -l}^{l} k \left( \frac{h}{H} \right) \Gamma_{h}
\end{equation*}
where $l = \min(H, J - 1)$ and $H$ is:
\begin{equation*}
H = \frac{1}{M}\sum_{i=1}^{M}  3.51 \cdot J^{3/5} \left( \frac{ (2 J)^{-1} \sum_{j = 1}^{J} r_{i,j}^{2}}{ \sum_{j = 1}^{\widetilde{J}} \widetilde{r}_{i,j}^{2} } \right)^{2/5}
\end{equation*}
with $r_{i,j}$ are the $i$-th elements of the vectors $r_{j}$. Similarly, $\widetilde{r}_{i,j}$ are the $i$-th elements of the vectors $\widetilde{r}_{j}$ where $\{\widetilde{r}_{j}\}_{j=1}^{\widetilde{J}}$ is the collection of return-vectors in the $j$-th bin of equally spaced $15$ minute intervals. $\Gamma_{h}$ and the Parzen kernel $k(x)$ are given by:
\begin{displaymath}
\Gamma_{h} = \left\{ 
\begin{array}{cc}
\displaystyle \sum_{j = h+1}^{J} r_{j} r^{\prime}_{j - h}  & \text{ if } h \geq 0 \\
\displaystyle \sum_{j = -h+1}^{J} r_{j+h} r^{\prime}_{j}   & \text{ if } h < 0 
\end{array}
\right.\qquad ; \qquad
k(x) =
\left\{
\begin{array}{ll}
1 - 6 x^{2} + 6 x^{3} & \text{if } x \in [0, 1/2] \\
2 (1 - x)^{3}         & \text{if } x \in (1/2, 1] \\
0                     & \text{otherwise}
\end{array}
\right.
\end{displaymath}

\subsection{Turnover Approximation\label{TOA}}

For our {\it day trader} type of investor, who opens (closes) all positions at the beginning (end) of the trading day, the average turnover of forecasting model $\kappa$ over $T$ trading days is given by:
\begin{equation*}
2\mathsf{TO}_{\kappa} = \frac{1}{T}\sum_{t=1}^{T}\sum_{j=1}^{M}\left|\widehat{\omega}_{\kappa,j,t}\right|+\left|\widehat{\omega}_{\kappa,j,t}^{c}\right|
\end{equation*}
where $\widehat{\omega}_{\kappa,j,t}^{c}$ is the value of the portfolio weight $\widehat{\omega}_{\kappa,j,t}$ at the end of the trading day. Value of the weights at {\it close} is related to value at {\it open} by $\widehat{\omega}_{\kappa,j,t}^{c}=(1+r_{j,t}^{oc})\cdot\widehat{\omega}_{\kappa,j,t}$, where $r_{j,t}^{oc}$ is the {\it open-close} return. It follows that turnover may be rewritten as:
\begin{equation*}
2\mathsf{TO}_{\kappa} = \frac{1}{T}\sum_{t=1}^{T}\sum_{j=1}^{M}(2+r_{j,t}^{oc})\left|\widehat{\omega}_{\kappa,j,t}\right|
\end{equation*}
Let $\overline{\overline{r}}$ be the daily weighted average return over the entire time series and across all assets:
\begin{equation*}
\overline{\overline{r}}\equiv \frac{\frac{1}{T\cdot M}\sum_{t=1}^{T}\sum_{j=1}^{M}|\widehat{\omega}_{\kappa,j,t}|\cdot r_{j,t}^{~oc}}{\frac{1}{T\cdot M}\sum_{t=1}^{T}\sum_{j=1}^{M}|\widehat{\omega}_{\kappa,j,t}|}
\end{equation*}
Then:
\begin{equation*}
\frac{1}{T}\sum_{t=1}^{T}\sum_{j=1}^{M}r_{j,t}^{~oc}|\widehat{\omega}_{\kappa,j,t}| = \overline{\overline{r}}\cdot\frac{1}{T}\sum_{t=1}^{T}\sum_{j=1}^{M}|\widehat{\omega}_{\kappa,j,t}|\\
\end{equation*}
from which it follows that turnover associated with model $\kappa$ may be rewritten as:
\begin{eqnarray*}
2\mathsf{TO}_{\kappa} & = & (2+\overline{\overline{r}})\cdot\frac{1}{T}\sum_{t=1}^{T}\sum_{j=1}^{M}\left|\widehat{\omega}_{\kappa,j,t}\right|\\
& \approx & 2\cdot\frac{1}{T}\sum_{t=1}^{T}\sum_{j=1}^{M}\left|\widehat{\omega}_{\kappa,j,t}\right|
\end{eqnarray*}
given that $\overline{\overline{r}}$ is usually a very small number. The main advantage of this approximation is not to rely on portfolio returns: since none of the competing models is optimized with respect to returns, including them in the measures of performance would only amount to adding noise to the analysis. In addition, to have an idea of the goodness of our approximation, consider the case in which the weighted average return is $p$ over a year's time, which corresponds to the daily average $\overline{\overline{r}}\approx 0.004\cdot p$. It follows that the percentage approximation error $\xi$ of our turnover measure is $\xi = -p(500+p)^{-1}$. Thus, even in the presence of a $100\%$ annual return, the percentage error of our measure of turnover is less than $0.2\%$.

\subsection{Transaction Costs Approximation\label{TCA}}

The approximation of average transaction cost associated with model $\kappa$ follows directly from that of turnover:
\begin{eqnarray*}
\overline{TC}_{\kappa} & = & \tau\cdot2\mathsf{TO}_{\kappa}\\
& \approx & 2\tau\cdot\frac{1}{T}\sum_{t=1}^{T}\sum_{j=1}^{M}\left|\widehat{\omega}_{\kappa,j,t}\right|
\end{eqnarray*}

\newpage

\begin{table}[ht]
\begin{center} 
\captionsetup{singlelinecheck=off}
\caption [ac] {

\scriptsize{Summary statistics of diagonal $\mathsf{DCW}$ parameter estimates over the five IS periods. The, equation by equation, sum of parameters $A^{*}+B^{*}$ captures persistence.}\label{tab:dcw_summary_stats}}
\resizebox{\textwidth}{!}{
\begin{tabular}{|clc| c r@{.}l r@{.}l r@{.}l ||clc| c r@{.}l r@{.}l r@{.}l r|}\toprule\toprule 
&&&& \multicolumn{2}{c}{$A^{*}$} & \multicolumn{2}{c}{$B^{*}$} & \multicolumn{2}{c}{$A^{*}+B^{*}$} & \multicolumn{4}{c}{} & \multicolumn{2}{c}{$A^{*}$} & \multicolumn{2}{c}{$B^{*}$} & \multicolumn{2}{c}{$A^{*}+B^{*}$} & \\
\toprule \toprule
& Mean       &&& 0&16585 & 0&77946 & 0&94531 &&  5\% perc. &&& 0&10437 & 0&60701 & 0&85147 & \\ 
& Minimum    &&& 0&07146 & 0&49990 & 0&77485 && 95\% perc. &&& 0&24781 & 0&85961 & 0&98286 & \\
& Maximum    &&& 0&35466 & 0&89259 & 0&98924 && Std.\ Dev. &&& 0&04287 & 0&07248 & 0&04051 & \\  
\bottomrule\bottomrule
\end{tabular}
}
%\end{center}
%\end{table}

\vspace{1cm}

%\begin{table}[ht]
%\begin{center} 
\captionsetup{singlelinecheck=off}
\caption [ac] {

\scriptsize{Summary statistics of IS $R^{2}$ for the diagonal $\mathsf{DCW}$ specification.}\label{tab:dcw_summary_stats_r2}}
\resizebox{\textwidth}{!}{
\begin{tabular}{|clc| c r@{.}l r@{.}l r@{.}l r@{.}l r@{.}l r@{.}l r|}\toprule\toprule 
&&&& \multicolumn{2}{c}{2005-09} & \multicolumn{2}{c}{2006-10} & \multicolumn{2}{c}{2007-11} & \multicolumn{2}{c}{2008-12} & \multicolumn{2}{c}{2009-13} & \multicolumn{2}{c}{2010-14} & \\
\toprule \toprule
& Mean       &&& 0&26927 & 0&28780 & 0&28042 & 0&23246 & 0&21310 & 0&20822 & \\
& Minimum    &&& 0&08605 & 0&09961 & 0&07387 & 0&07048 & 0&04422 & 0&06710 & \\
& Maximum    &&& 0&46056 & 0&49641 & 0&51059 & 0&48423 & 0&43555 & 0&41336 & \\
&  5\% perc. &&& 0&09369 & 0&11291 & 0&11389 & 0&07357 & 0&05967 & 0&07272 & \\
& 95\% perc. &&& 0&44997 & 0&48114 & 0&48190 & 0&44085 & 0&41778 & 0&39195 & \\
& Std.\ Dev. &&& 0&10319 & 0&11272 & 0&09711 & 0&09085 & 0&09794 & 0&09212 & \\
\bottomrule\bottomrule
\end{tabular}
}
%\end{center}
%\end{table}

\vspace{1cm}

%\begin{table}[ht]
%\begin{center} 
\captionsetup{singlelinecheck=off}
\caption [ac] {
\scriptsize{Summary statistics of OOS $R^{2}$ for the diagonal $\mathsf{DCW}$ specification. The OOS total sums of squares are calculated from the OOS averages.}\label{tab:dcw_summary_stats_r2_out}}
\resizebox{\textwidth}{!}{
\begin{tabular}{|clc| c r@{.}l r@{.}l r@{.}l r@{.}l r@{.}l r@{.}l r|}\toprule\toprule 
&&&& \multicolumn{2}{c}{2010} & \multicolumn{2}{c}{2011} & \multicolumn{2}{c}{2012} & \multicolumn{2}{c}{2013} & \multicolumn{2}{c}{2014} & \multicolumn{2}{c}{2015} & \\
\toprule \toprule
& Mean       &&& 0&12704 & 0&12773 & 0&08409 & 0&05893 & 0&10469 & 0&12473 & \\
& Minimum    &&&-0&06260 &-0&02693 &-0&05123 &-0&04390 &-0&06979 &-0&03695 & \\
& Maximum    &&& 0&33471 & 0&40899 & 0&37336 & 0&21401 & 0&25642 & 0&36444 & \\
& Std.\ Dev. &&& 0&09520 & 0&10447 & 0&09785 & 0&06753 & 0&08564 & 0&09834 & \\
&  5\% perc. &&&-0&03601 &-0&01474 &-0&04148 &-0&04130 &-0&05593 &-0&02137 & \\
& 95\% perc. &&& 0&32166 & 0&36616 & 0&31808 & 0&20497 & 0&24894 & 0&33456 & \\
\bottomrule\bottomrule
\end{tabular}
}
\end{center}
\end{table}

\begin{table}[ht]
\begin{center} 
\captionsetup{singlelinecheck=off}
\caption [ac] {

\scriptsize{Average OOS daily variances $\mathsf{PV}$ of the portfolio strategies. For $\mathsf{Naive}$ and $\mathsf{VT}$, exposure $EC$ is $1.00$ by definition. For $\mathsf{RW}$, $\mathsf{DCC}$ and $\mathsf{DCW}$ results are presented without exposure constraints ($EC\le\infty$), with $EC\le\{2.00,1.75,1.50,1.25\}$ and no short-selling ($EC=1.00$).}\label{tab:table01}}
\resizebox{\textwidth}{!}{
\begin{tabular}{|clc| c r@{.}l r@{.}l r@{.}l r@{.}l r@{.}l r@{.}l r@{.}l r|}\toprule\toprule 
& \textsc{Model} &&& \multicolumn{2}{r}{2010\textcolor{white}{$^{***}$}} & \multicolumn{2}{r}{2011\textcolor{white}{$^{***}$}} & \multicolumn{2}{r}{2012\textcolor{white}{$^{***}$}} & \multicolumn{2}{r}{2013\textcolor{white}{$^{***}$}} & \multicolumn{2}{r}{2014\textcolor{white}{$^{***}$}} & \multicolumn{2}{r}{2015\textcolor{white}{$^{***}$}} & \multicolumn{2}{r}{\textsc{All}\textcolor{white}{$^{***}$}} & \\
\toprule \toprule
& $\mathsf{Naive}$  &&& 0&765768 & 0&940376 & 0&337162 & 0&247769 & 0&265182 & 0&477073 & 0&505796 &\\
& $\mathsf{VT}$     &&& 0&629024 & 0&776975 & 0&277726 & 0&228755 & 0&242804 & 0&443216 & 0&433283 &\\
\midrule
\multicolumn{19}{|c|}{$EC\le\infty$}\\
\midrule
& $\mathsf{RW}$     &&& 0&410682 & 0&440253 & 0&220276 & 0&249329 & 0&223546 & 0&415370 & 0&326658 &\\
& $\mathsf{DCC}$    &&& 0&363172 & 0&424437 & 0&199700 & 0&211095 & 0&206934 & 0&376261 & 0&297010 &\\
& $\mathsf{DCW}$    &&& 0&368602 & 0&404762 & 0&180981 & 0&200283 & 0&190229 & 0&353792 & 0&283197 &\\
\midrule
\multicolumn{19}{|c|}{$EC\le 2.00$}\\
\midrule
& $\mathsf{RW}$     &&& 0&398498 & 0&429203 & 0&215218 & 0&241141 & 0&216777 & 0&386768 & 0&314685 &\\
& $\mathsf{DCC}$    &&& 0&363755 & 0&424242 & 0&199700 & 0&211045 & 0&206517 & 0&375678 & 0&296899 &\\
& $\mathsf{DCW}$    &&& 0&373493 & 0&418564 & 0&181196 & 0&199287 & 0&190273 & 0&359317 & 0&287114 &\\
\midrule
\multicolumn{19}{|c|}{$EC\le 1.75$}\\
\midrule
& $\mathsf{RW}$     &&& 0&389666 & 0&431561 & 0&209392 & 0&235116 & 0&211803 & 0&381235 & 0&309880 &\\
& $\mathsf{DCC}$    &&& 0&365464 & 0&429841 & 0&199629 & 0&210688 & 0&205754 & 0&374272 & 0&297687 &\\
& $\mathsf{DCW}$    &&& 0&379125 & 0&428851 & 0&181590 & 0&198752 & 0&190562 & 0&362648 & 0&290351 &\\
\midrule
\multicolumn{19}{|c|}{$EC\le 1.50$}\\
\midrule
& $\mathsf{RW}$     &&& 0&387697 & 0&441894 & 0&202447 & 0&227201 & 0&206624 & 0&378370 & 0&307464 &\\
& $\mathsf{DCC}$    &&& 0&370794 & 0&442001 & 0&198367 & 0&209836 & 0&204315 & 0&372569 & 0&299733 &\\
& $\mathsf{DCW}$    &&& 0&390087 & 0&445715 & 0&183172 & 0&198466 & 0&191394 & 0&368115 & 0&296260 &\\
\midrule
\multicolumn{19}{|c|}{$EC\le 1.25$}\\
\midrule
& $\mathsf{RW}$     &&& 0&396078 & 0&466123 & 0&198625 & 0&218991 & 0&202397 & 0&379484 & 0&310385 &\\
& $\mathsf{DCC}$    &&& 0&385483 & 0&467761 & 0&196376 & 0&207064 & 0&202168 & 0&374460 & 0&305651 &\\
& $\mathsf{DCW}$    &&& 0&410382 & 0&478605 & 0&187242 & 0&199859 & 0&193726 & 0&376030 & 0&307905 &\\
\midrule
\multicolumn{19}{|c|}{$EC=1.00$}\\
\midrule
& $\mathsf{RW}$     &&& 0&416423 & 0&509316 & 0&206378 & 0&223728 & 0&205606 & 0&391573 & 0&325618 &\\
& $\mathsf{DCC}$    &&& 0&399477 & 0&510957 & 0&201358 & 0&208639 & 0&201557 & 0&383752 & 0&317734 &\\
& $\mathsf{DCW}$    &&& 0&435044 & 0&516830 & 0&193881 & 0&205725 & 0&196801 & 0&386463 & 0&322585 &\\
\bottomrule\bottomrule
\end{tabular}
}
\end{center}
\end{table}

\begin{table}[ht]
\begin{center} 
\captionsetup{singlelinecheck=off}
\caption [ac] {

\scriptsize{Average OOS daily certainty equivalent $\mathsf{CEQ}$, expressed in {\it basis points}, relative to the change of strategy indicated by $\rightarrow$. $\mathsf{CEQ}$ are calculated for a risk-aversion coefficient of $\gamma=1$ and may be computed for different values of $\gamma$ by simple multiplication. For $\mathsf{Naive}$ and $\mathsf{VT}$, exposure $EC$ is $1.00$ by definition. For $\mathsf{RW}$, $\mathsf{DCC}$ and $\mathsf{DCW}$ results are presented without exposure constraints ($EC\le\infty$), with $EC\le\{2.00,1.75,1.50,1.25\}$ and no short-selling ($EC=1.00$).}\label{tab:table02}}
\resizebox{\textwidth}{!}{
\begin{tabular}{|clc| c r@{.}l r@{.}l r@{.}l r@{.}l r@{.}l r@{.}l r@{.}l r|}\toprule\toprule 
& \textsc{Model} &&& \multicolumn{2}{r}{2010\textcolor{white}{$^{***}$}} & \multicolumn{2}{r}{2011\textcolor{white}{$^{***}$}} & \multicolumn{2}{r}{2012\textcolor{white}{$^{***}$}} & \multicolumn{2}{r}{2013\textcolor{white}{$^{***}$}} & \multicolumn{2}{r}{2014\textcolor{white}{$^{***}$}} & \multicolumn{2}{r}{2015\textcolor{white}{$^{***}$}} & \multicolumn{2}{r}{\textsc{All}\textcolor{white}{$^{***}$}} & \\
\toprule \toprule
& $\mathsf{Naive}\rightarrow\mathsf{VT}$   &&& ~~~~6&84~~~ & ~~~~8&17~~~ & ~~~~2&97~~~ & ~~~~0&95~~~ & ~~~~1&12~~~ & ~~~~1&69~~~ & ~~~~3&63~~~ &\\
\midrule
\multicolumn{19}{|c|}{$EC\le\infty$}\\
\midrule
& $\mathsf{VT}\rightarrow\mathsf{RW}$      &&& 10&92 & 16&84 & 2&87 & -1&03 & 0&96 & 1&39 & 5&33 &\\
& $\mathsf{RW}\rightarrow\mathsf{DCC}$     &&&  2&38 & 0&79 & 1&03 & 1&91 & 0&83 & 1&96 & 1&48 &\\
& $\mathsf{DCC}\rightarrow\mathsf{DCW}$    &&& -0&27 & 0&98 & 0&94 & 0&54 & 0&84 & 1&12 & 0&69 &\\
\midrule
\multicolumn{19}{|c|}{$EC\le 2.00$}\\
\midrule
& $\mathsf{VT}\rightarrow\mathsf{RW}$      &&& 11&53 & 17&39 & 3&13 & -0&62 & 1&30 & 2&82 & 5&93 &\\
& $\mathsf{RW}\rightarrow\mathsf{DCC}$     &&&  1&74 & 0&25 & 0&78 & 1&50 & 0&51 & 0&55 & 0&89 &\\
& $\mathsf{DCC}\rightarrow\mathsf{DCW}$    &&& -0&49 & 0&28 & 0&93 & 0&59 & 0&81 & 0&82 & 0&49 &\\
\midrule
\multicolumn{19}{|c|}{$EC\le 1.75$}\\
\midrule
& $\mathsf{VT}\rightarrow\mathsf{RW}$      &&& 11&97 & 17&27 & 3&42 & -0&32 & 1&55 & 3&10 & 6&17 &\\
& $\mathsf{RW}\rightarrow\mathsf{DCC}$     &&&  1&21 & 0&09 & 0&49 & 1&22 & 0&30 & 0&35 & 0&61 &\\
& $\mathsf{DCC}\rightarrow\mathsf{DCW}$    &&& -0&68 & 0&05 & 0&90 & 0&60 & 0&76 & 0&58 & 0&37 &\\
\midrule
\multicolumn{19}{|c|}{$EC\le 1.50$}\\
\midrule
& $\mathsf{VT}\rightarrow\mathsf{RW}$      &&& 12&07 & 16&75 & 3&76 & 0&08 & 1&81 & 3&24 & 6&29 &\\
& $\mathsf{RW}\rightarrow\mathsf{DCC}$     &&& 0&85 & -0&01 & 0&20 & 0&87 & 0&12 & 0&29 & 0&39 &\\
& $\mathsf{DCC}\rightarrow\mathsf{DCW}$    &&& -0&96 & -0&19 & 0&76 & 0&57 & 0&65 & 0&22 & 0&17 &\\
\midrule
\multicolumn{19}{|c|}{$EC\le 1.25$}\\
\midrule
& $\mathsf{VT}\rightarrow\mathsf{RW}$      &&& 11&65 & 15&54 & 3&96 & 0&49 & 2&02 & 3&19 & 6&14 &\\
& $\mathsf{RW}\rightarrow\mathsf{DCC}$     &&& 0&53 & -0&08 & 0&11 & 0&60 & 0&01 & 0&25 & 0&24 &\\
& $\mathsf{DCC}\rightarrow\mathsf{DCW}$    &&& -1&24 & -0&54 & 0&46 & 0&36 & 0&42 & -0&08 & -0&11 &\\
\midrule
\multicolumn{19}{|c|}{$EC= 1.00$}\\
\midrule
& $\mathsf{VT}\rightarrow\mathsf{RW}$      &&& 10&63 & 13&38 & 3&57 & 0&25 & 1&86 & 2&58 & 5&38 &\\
& $\mathsf{RW}\rightarrow\mathsf{DCC}$     &&& 0&85 & -0&08 & 0&25 & 0&75 & 0&20 & 0&39 & 0&39 &\\
& $\mathsf{DCC}\rightarrow\mathsf{DCW}$    &&& -1&78 & -0&29 & 0&37 & 0&15 & 0&24 & -0&14 & -0&24 &\\
\bottomrule\bottomrule
\end{tabular}
}
\end{center}
\end{table}

\begin{table}[ht]
\begin{center} 
\captionsetup{singlelinecheck=off}
\caption [ac] {

\scriptsize{Average daily OOS daily turnover $\mathsf{TO}$ of the portfolio strategies. For $\mathsf{Naive}$ and $\mathsf{VT}$, exposure $EC$ is $1.00$ by definition. For $\mathsf{RW}$, $\mathsf{DCC}$ and $\mathsf{DCW}$ results are presented without exposure constraints ($EC\le\infty$), with $EC\le\{2.00,1.75,1.50,1.25\}$ and no short-selling ($EC=1.00$).}\label{tab:table03}}
\resizebox{\textwidth}{!}{
\begin{tabular}{|clc| c r@{.}l r@{.}l r@{.}l r@{.}l r@{.}l r@{.}l r@{.}l r|}\toprule\toprule 
& \textsc{Model} &&& \multicolumn{2}{r}{2010\textcolor{white}{$^{***}$}} & \multicolumn{2}{r}{2011\textcolor{white}{$^{***}$}} & \multicolumn{2}{r}{2012\textcolor{white}{$^{***}$}} & \multicolumn{2}{r}{2013\textcolor{white}{$^{***}$}} & \multicolumn{2}{r}{2014\textcolor{white}{$^{***}$}} & \multicolumn{2}{r}{2015\textcolor{white}{$^{***}$}} & \multicolumn{2}{r}{\textsc{All}\textcolor{white}{$^{***}$}} & \\
\toprule \toprule
& $\mathsf{Naive}$  &&& ~~~~1&00~~~ & ~~~~1&00~~~ & ~~~~1&00~~~ & ~~~~1&00~~~ & ~~~~1&00~~~ & ~~~~1&00~~~ & ~~~~1&00~~~ &\\
& $\mathsf{VT}$     &&& 1&00 & 1&00 & 1&00 & 1&00 & 1&00 & 1&00 & 1&00 &\\
\midrule
\multicolumn{19}{|c|}{$EC\le\infty$}\\
\midrule
& $\mathsf{RW}$     &&& 2&06 & 2&09 & 1&84 & 1&83 & 1&80 & 2&17 & 1&97 &\\
& $\mathsf{DCC}$    &&& 1&68 & 1&74 & 1&49 & 1&43 & 1&47 & 1&69 & 1&59 &\\
& $\mathsf{DCW}$    &&& 1&49 & 1&55 & 1&34 & 1&27 & 1&21 & 1&42 & 1&38 &\\
\midrule
\multicolumn{19}{|c|}{$EC\le 2.00$}\\
\midrule
& $\mathsf{RW}$     &&& 1&87 & 1&84 & 1&77 & 1&73 & 1&70 & 1&83 & 1&79 &\\
& $\mathsf{DCC}$    &&& 1&67 & 1&69 & 1&49 & 1&43 & 1&46 & 1&66 & 1&57 &\\
& $\mathsf{DCW}$    &&& 1&42 & 1&45 & 1&32 & 1&23 & 1&18 & 1&29 & 1&32 &\\
\midrule
\multicolumn{19}{|c|}{$EC \le 1.75$}\\
\midrule
& $\mathsf{RW}$     &&& 1&71 & 1&69 & 1&67 & 1&62 & 1&60 & 1&67 & 1&66 &\\
& $\mathsf{DCC}$    &&& 1&61 & 1&61 & 1&49 & 1&42 & 1&44 & 1&60 & 1&53 &\\
& $\mathsf{DCW}$    &&& 1&36 & 1&39 & 1&28 & 1&20 & 1&15 & 1&23 & 1&27 &\\
\midrule
\multicolumn{19}{|c|}{$EC\le 1.50$}\\
\midrule
& $\mathsf{RW}$     &&& 1&49 & 1&49 & 1&49 & 1&46 & 1&43 & 1&46 & 1&47 &\\
& $\mathsf{DCC}$    &&& 1&47 & 1&46 & 1&43 & 1&38 & 1&38 & 1&45 & 1&43 &\\
& $\mathsf{DCW}$    &&& 1&26 & 1&29 & 1&22 & 1&15 & 1&11 & 1&14 & 1&19 &\\
\midrule
\multicolumn{19}{|c|}{$EC\le 1.25$}\\
\midrule
& $\mathsf{RW}$     &&& 1&25 & 1&25 & 1&25 & 1&22 & 1&22 & 1&22 & 1&23 &\\
& $\mathsf{DCC}$    &&& 1&25 & 1&25 & 1&25 & 1&22 & 1&20 & 1&22 & 1&23 &\\
& $\mathsf{DCW}$    &&& 1&13 & 1&15 & 1&12 & 1&07 & 1&05 & 1&05 & 1&09 &\\
\midrule
\multicolumn{19}{|c|}{$EC=1.00$}\\
\midrule
& $\mathsf{RW}$     &&& 1&00 & 1&00 & 1&00 & 1&00 & 1&00 & 1&00 & 1&00 &\\
& $\mathsf{DCC}$    &&& 1&00 & 1&00 & 1&00 & 1&00 & 1&00 & 1&00 & 1&00 &\\
& $\mathsf{DCW}$    &&& 1&00 & 1&00 & 1&00 & 1&00 & 1&00 & 1&00 & 1&00 &\\
\bottomrule\bottomrule
\end{tabular}
}
\end{center}
\end{table}

\begin{table}[ht]
\begin{center} 
\captionsetup{singlelinecheck=off}
\caption [ac] {

\scriptsize{Average OOS daily break-even transaction costs $\mathsf{BETC}$, expressed in {\it basis points}, relative to the change of strategy indicated by $\rightarrow$. $\mathsf{BETC}$ are calculated for a risk-aversion coefficient of $\gamma=1$ and may be computed for different values of $\gamma$ by simple multiplication. $<$ and $>$ define the range of transaction costs for which the strategy on the right of $\rightarrow$ is preferred to that on the left. The entry A (N) indicates that the strategy to the right of $\rightarrow$ is preferred for Any (No) value of $\tau/\gamma$. For $\mathsf{Naive}$ and $\mathsf{VT}$, exposure $EC$ is $1.00$ by definition. For $\mathsf{RW}$, $\mathsf{DCC}$ and $\mathsf{DCW}$ results are presented without exposure constraints ($EC\le\infty$), with $EC\le\{2.00,1.75,1.50,1.25\}$ and no short-selling ($EC=1.00$).}\label{tab:table04}}
\resizebox{\textwidth}{!}{
\begin{tabular}{|clc| c r@{.}l r@{.}l r@{.}l r@{.}l r@{.}l r@{.}l r@{.}l r|}\toprule\toprule 
& \textsc{Model} &&& \multicolumn{2}{r}{2010\textcolor{white}{$^{***}$}} & \multicolumn{2}{r}{2011\textcolor{white}{$^{***}$}} & \multicolumn{2}{r}{2012\textcolor{white}{$^{***}$}} & \multicolumn{2}{r}{2013\textcolor{white}{$^{***}$}} & \multicolumn{2}{r}{2014\textcolor{white}{$^{***}$}} & \multicolumn{2}{r}{2015\textcolor{white}{$^{***}$}} & \multicolumn{2}{r}{\textsc{All}\textcolor{white}{$^{***}$}} & \\
\toprule \toprule 
& $\mathsf{Naive}\rightarrow\mathsf{VT}$   &&& \multicolumn{2}{c}{A} & \multicolumn{2}{c}{A} & \multicolumn{2}{c}{A} & \multicolumn{2}{c}{A} & \multicolumn{2}{c}{A} & \multicolumn{2}{c}{A} & \multicolumn{2}{c}{A} &\\
\midrule
\multicolumn{19}{|c|}{$EC\le\infty$}\\
\midrule
& $\mathsf{VT}\rightarrow\mathsf{RW}$      &&& ~~~$<$5&15~~~ & ~~~$<$7&72~~~ & ~~~$<$1&71~~~ & \multicolumn{2}{c}{N} & ~~~$<$0&60~~~ & ~~~$<$0&60~~~ & ~~~$<$2&75~~~ & \\
& $\mathsf{RW}\rightarrow\mathsf{DCC}$     &&& \multicolumn{2}{c}{A} & \multicolumn{2}{c}{A} & \multicolumn{2}{c}{A} & \multicolumn{2}{c}{A} & \multicolumn{2}{c}{A} & \multicolumn{2}{c}{A} & \multicolumn{2}{c}{A} & \\
& $\mathsf{DCC}\rightarrow\mathsf{DCW}$    &&& $<$0&71 & \multicolumn{2}{c}{A} & \multicolumn{2}{c}{A} & \multicolumn{2}{c}{A} & \multicolumn{2}{c}{A} & \multicolumn{2}{c}{A} & \multicolumn{2}{c}{A} & \\
\midrule
\multicolumn{19}{|c|}{$EC\le 2.00$}\\
\midrule
& $\mathsf{VT}\rightarrow\mathsf{RW}$      &&& $<$6&62 & $<$10&35 & $<$2&03 & \multicolumn{2}{c}{N} & $<$0&93 & $<$1&70 & $<$3&75 & \\
& $\mathsf{RW}\rightarrow\mathsf{DCC}$     &&& \multicolumn{2}{c}{A} & \multicolumn{2}{c}{A} & \multicolumn{2}{c}{A} & \multicolumn{2}{c}{A} & \multicolumn{2}{c}{A} & \multicolumn{2}{c}{A} & \multicolumn{2}{c}{A} & \\
& $\mathsf{DCC}\rightarrow\mathsf{DCW}$    &&& \multicolumn{2}{c}{A} & \multicolumn{2}{c}{A} & \multicolumn{2}{c}{A} & \multicolumn{2}{c}{A} & \multicolumn{2}{c}{A} & \multicolumn{2}{c}{A} & \multicolumn{2}{c}{A} & \\
\midrule
\multicolumn{19}{|c|}{$EC\le 1.75$}\\
\midrule
& $\mathsf{VT}\rightarrow\mathsf{RW}$      &&& $<$8&43 & $<$12&53 & $<$2&55 & \multicolumn{2}{c}{N} & $<$1&29 & $<$2&31 & $<$4&67 &\\
& $\mathsf{RW}\rightarrow\mathsf{DCC}$     &&& \multicolumn{2}{c}{A} & \multicolumn{2}{c}{A} & \multicolumn{2}{c}{A} & \multicolumn{2}{c}{A} & \multicolumn{2}{c}{A} & \multicolumn{2}{c}{A} & \multicolumn{2}{c}{A} & \\
& $\mathsf{DCC}\rightarrow\mathsf{DCW}$    &&& $>$1&37 & \multicolumn{2}{c}{A} & \multicolumn{2}{c}{A} & \multicolumn{2}{c}{A} & \multicolumn{2}{c}{A} & \multicolumn{2}{c}{A} & \multicolumn{2}{c}{A} & \\
\midrule
\multicolumn{19}{|c|}{$EC\le 1.50$}\\
\midrule
& $\mathsf{VT}\rightarrow\mathsf{RW}$      &&& $<$12&31 & $<$17&10 & $<$3&84 & \multicolumn{2}{c}{N} & $<$2&10 & $<$3&52 & $<$6&69 &\\
& $\mathsf{RW}\rightarrow\mathsf{DCC}$     &&& \multicolumn{2}{c}{A} & \multicolumn{2}{c}{A} & \multicolumn{2}{c}{A} & \multicolumn{2}{c}{A} & \multicolumn{2}{c}{A} & \multicolumn{2}{c}{A} & \multicolumn{2}{c}{A} & \\
& $\mathsf{DCC}\rightarrow\mathsf{DCW}$    &&& $>$2&30 & $>$0&55 & \multicolumn{2}{c}{A} & \multicolumn{2}{c}{A} & \multicolumn{2}{c}{A} & \multicolumn{2}{c}{A} & \multicolumn{2}{c}{A} & \\
\midrule
\multicolumn{19}{|c|}{$EC\le 1.25$}\\
\midrule
& $\mathsf{VT}\rightarrow\mathsf{RW}$      &&& $<$23&29 & $<$31&09 & $<$7&91 & ~~~$<$1&11~~~ & $<$4&59 & $<$7&24 & $<$13&36 &\\
& $\mathsf{RW}\rightarrow\mathsf{DCC}$     &&& \multicolumn{2}{c}{A} & \multicolumn{2}{c}{N} & \multicolumn{2}{c}{A} & \multicolumn{2}{c}{A} & \multicolumn{2}{c}{A} & \multicolumn{2}{c}{A} & \multicolumn{2}{c}{A} & \\
& $\mathsf{DCC}\rightarrow\mathsf{DCW}$    &&& $>$5&19 & $>$2&71 & \multicolumn{2}{c}{A} & \multicolumn{2}{c}{A} & \multicolumn{2}{c}{A} & $>$0&23 & $>$0&40 & \\
\midrule
\multicolumn{19}{|c|}{$EC=1.00$}\\
\midrule
& $\mathsf{VT}\rightarrow\mathsf{RW}$      &&& \multicolumn{2}{c}{A} & \multicolumn{2}{c}{A} & \multicolumn{2}{c}{A} & \multicolumn{2}{c}{A} & \multicolumn{2}{c}{A} & \multicolumn{2}{c}{A} & \multicolumn{2}{c}{A} &\\
& $\mathsf{RW}\rightarrow\mathsf{DCC}$     &&& \multicolumn{2}{c}{A} & \multicolumn{2}{c}{N} & \multicolumn{2}{c}{A} & \multicolumn{2}{c}{A} & \multicolumn{2}{c}{A} & \multicolumn{2}{c}{A} & \multicolumn{2}{c}{A} & \\
& $\mathsf{DCC}\rightarrow\mathsf{DCW}$    &&& \multicolumn{2}{c}{N} & \multicolumn{2}{c}{N} & \multicolumn{2}{c}{A} & \multicolumn{2}{c}{A} & \multicolumn{2}{c}{A} & \multicolumn{2}{c}{N} & \multicolumn{2}{c}{N} & \\
\bottomrule\bottomrule
\end{tabular}
}
\end{center}
\end{table}

\newpage

\newpage

\newpage

\end{document}